\documentclass[12pt,a4paper]{article}
\usepackage{amsmath,amsfonts}
\usepackage{algorithm}
\usepackage{array}
\usepackage[caption=false,font=footnotesize]{subfig}
\usepackage{textcomp}
\usepackage{stfloats}
\usepackage{url}
\usepackage{verbatim}
\usepackage{graphicx}
\usepackage{cite}
\usepackage{hyperref}
\usepackage{algpseudocode}
\usepackage{xcolor}
\usepackage{bm}
\usepackage{multirow}
\usepackage{booktabs}
\usepackage{threeparttable}
\usepackage{geometry}
\usepackage{setspace}
\begin{document}

\title{\textbf{Active Learning for Channel Knowledge Map Construction via Bayesian Inference Diffusion Models}}

\author{Yunzhe~Zhu, Xuewen~Liao, Zhenzhen~Gao, Linzhou~Zeng, Yong~Zeng,}






\maketitle

\begin{abstract}
  Channel knowledge maps (CKMs) are regarded as key enablers of environment-aware communications in future wireless networks, as they provide location-specific channel information by establishing an explicit connection between wireless devices and the physical propagation environment. As a representative CKM, the channel gain map (CGM) characterizes the spatial distributions of large-scale fading to support wireless environment awareness and network optimization. Existing CGM construction methods generally lack a well-defined sampling-point acquisition strategy, which may result in a limited number of sampling points being allocated to spatially redundant or highly predictable regions, thereby degrading CGM reconstruction performance in complex propagation environments. In this paper, we propose an active-learning-based diffusion framework for efficient CGM construction. By combining Bayesian inference with the diffusion model, the proposed method estimates epistemic uncertainty without retraining the model. Two uncertainty quantification algorithms are further developed along the reverse diffusion process to generate element-wise epistemic uncertainty maps. Furthermore, an uncertainty-aware sampling strategy is designed to determine new observation locations by jointly considering epistemic uncertainty and spatial distribution uniformity. Experimental results on both static and dynamic CGM datasets demonstrate that the proposed method achieves better reconstruction performance than baseline methods. These results indicate that the proposed method can effectively improve the utilization efficiency of limited sampling points and enhance the accuracy of CGM construction in complex wireless propagation environments.
\end{abstract}


\section{Introduction}
Channel knowledge maps (CKMs) \cite{Zeng2021,Zeng2024} can serve as a bridge between wireless devices and the physical environment. With the aid of CKM, the conventional communication paradigm without environmental awareness can be transformed into an environment-aware communication paradigm, making CKM of significant research interest in the 6G era. A CKM can be viewed as a knowledge base that offers channel-related prior information associated with each location in space. Such information may include the line-of-sight/non-line-of-sight (LOS/NLOS) propagation state, path loss, shadow fading, as well as small-scale fading channel characteristics, such as multipath delay, Doppler power spectrum, and angular power spectrum. 
The reason why CKM can enable wireless devices to perceive the environment lies in the fact that wireless channel fading characteristics are primarily determined by the local propagation environment. For example, at the same carrier frequency, large-scale fading is mainly affected by the spatial distribution and material properties of obstacles. For small-scale fading, the number of multipath components, their powers, angles of arrival and departure, and delays are further determined by obstacles in the environment, while Doppler shifts are governed by the velocities of movable obstacles. Therefore, with the aid of CKM, wireless devices can shift from passively experiencing wireless channel fading and attempting to overcome or mitigate its adverse effects to perceiving the propagation environment through channel-related information and exploiting environmental characteristics to improve the performance of communication systems.

As one of the most fundamental and commonly used forms of CKM, a CGM aims to establish the mapping relationship between spatial locations and channel gains or path losses within a given target area. Specifically, for any transmitter-receiver location pair, a CGM can provide the location-specific average channel attenuation, which is typically represented by metrics such as path loss, received signal strength, or large-scale channel gain. Since path loss directly reflects the degree of signal power attenuation during wireless propagation, a CGM can intuitively characterize the wireless coverage quality, signal strength distribution, and potential coverage holes in a map-based manner. Therefore, CGM can be applied to indoor or urban canyon localization \cite{9031749,10025691}, path planning for unmanned aerial vehicles or mobile robots \cite{8525324,9269485,9354009}, base station deployment \cite{10008712}, user resource scheduling and link adaptation \cite{10474197}, physical-layer security \cite{8815467}, device-to-device (D2D) communications \cite{9053347}, and other related scenarios.

In complex wireless scenarios, it is often unrealistic to conduct point-by-point, dense, and repeated field measurements over the entire area. To reduce the measurement overhead, a more practically feasible solution is to collect sparse observations at only a limited number of locations and then recover or infer the channel gain distribution over the entire target region. From a mathematical perspective, this problem is essentially a typical inverse problem \cite{fu2025ckmdiff}, where a high-dimensional spatial field distribution needs to be recovered from limited observations. Since the number of observations is usually much smaller than the number of locations to be estimated, this problem is inherently underdetermined and ill-posed. 

Traditional CGM construction methods are mainly based on the idea of spatial interpolation. These methods usually assume that the channel gains at neighboring spatial locations exhibit a certain degree of continuity or correlation, and accordingly infer the channel states at unobserved locations from the measurements collected at observed locations. Representative methods include inverse distance weighting interpolation \cite{Lu2008}, radial basis function interpolation \cite{Bishop1995}, Kriging interpolation \cite{1371308}, and matrix completion \cite{7472907}. 
However, spatial interpolation methods based on conventional spatial smoothness assumptions usually cannot fully characterize the propagation features in complex environments, and their reconstruction accuracy and generalization capability are significantly limited.

Deep learning-based methods have been investigated as a data-driven alternative for CGM construction, since they can learn nonlinear mappings from sparse observations and environmental structures to complete channel gain distributions without imposing explicit spatial-correlation assumptions or manually designed kernels. Existing approaches can be broadly divided into discriminative and generative models. Discriminative methods directly estimate the missing CGM values conditioned on the available observations, and representative architectures include convolutional autoencoders \cite{9523765, 10682525}, U-Net-based networks \cite{9354041, 11146461}, ViT-based models that exploit self-attention to capture long-range spatial dependencies \cite{11174709}, and graph neural networks that characterize spatial correlations by modeling CGM locations as nodes and edges \cite{10682510, 10078269}. Although these methods improve reconstruction accuracy over conventional interpolation, their deterministic prediction mechanisms tend to produce overly smooth maps and may fail to recover high-frequency spatial details. To address this limitation, generative models have been introduced to learn the underlying distribution and structural priors of CGMs. GAN-based methods have been applied to CGM reconstruction \cite{8794603,10130091}, but they are often affected by training instability, convergence difficulties, and mode collapse. More recently, diffusion models have attracted increasing attention for CGM construction owing to their strong distribution modeling capability and iterative denoising mechanism \cite{10764739, 11278649, huang2026channel}.

However, most studies on CGM reconstruction from sparse observations assume randomly sampled observation locations. Since practical measurements are usually scarce, random sampling may allocate resources to information-redundant or easily predictable regions, thereby limiting the reconstruction gain obtained from new measurements. The influence of sampling locations has been partially studied. For example, \cite{10530520} theoretically analyzed the number of samples required for CGM construction with spatial interpolation, but considered only random and uniform sampling without addressing where the observations should be placed. Reference \cite{10735108} introduced model uncertainty to guide observation selection for CGM reconstruction, where Gaussian process regression was used and its predictive variance naturally quantified location-dependent uncertainty. Nevertheless, the limited reconstruction capability of Gaussian process regression in complex propagation environments restricts its practical applicability.

Therefore, the core idea of this paper is to introduce Bayesian inference into the diffusion model to estimate element-wise epistemic uncertainty of CGM reconstruction, which is referred to as the CGM prediction variance map in this paper. Epistemic uncertainty reflects the degree of predictive uncertainty caused by limited training data, insufficient observations, or high scene complexity. For a given spatial location, a large predictive uncertainty indicates that the model lacks sufficiently reliable prior knowledge or observational support at that location. Accordingly, these high-uncertainty regions usually contain more valuable information for CGM reconstruction. Therefore, preferentially selecting locations with high epistemic uncertainty as new sampling points can concentrate limited measurement resources on the regions where the model most needs additional information, thereby effectively reducing the overall reconstruction error.

Based on the above discussions, the main contributions of this paper can be summarized as follows:
\begin{enumerate}
\item{We design two algorithms to quantify the epistemic uncertainty of CGM without retraining the diffusion model. Specifically, the last-layer Laplace approximation is adopted to estimate the posterior distribution of model parameters. Based on this posterior distribution, we derive the formulation for propagating the posterior uncertainty of model parameters to the uncertainty of generated CGM samples during the reverse diffusion process. Accordingly, two specific implementation algorithms are developed. The first algorithm approximately computes the element-wise epistemic uncertainty of the generated results through a single reverse diffusion process. The second algorithm propagates parameter uncertainty through multiple reverse diffusion processes to accelerate variance estimation of the generated results.}
\item{Directly selecting the top-K locations with the largest uncertainty may concentrate samples in a few highly uncertain regions and produce redundant observations. To address this issue, we design an uncertainty-aware sampling algorithm that jointly considers epistemic uncertainty and spatial distribution uniformity. The proposed algorithm encourages the sampling points to be preferentially distributed in regions with large predictive variance, so as to acquire more informative observations, while preventing excessive concentration of sampling points through a spatial constraint mechanism.}
\item{We propose an active-learning-based diffusion framework for efficient CGM construction from sparse observations. The framework uses the diffusion model to reconstruct complete CGMs and integrates the uncertainty-aware sampling mechanism to select candidate locations with high epistemic uncertainty, thereby allocating limited measurement resources to regions most beneficial for improving reconstruction performance.}
\end{enumerate}

The rest of this paper is organized as follows. Section~\ref{PF} formulates the sparse-observation CGM reconstruction problem and reviews the required preliminaries, including diffusion models, uncertainty quantification, and the Laplace approximation. Section~\ref{EUQBDM} derives the two proposed diffusion-based epistemic uncertainty quantification methods. Section~\ref{UASA} presents the uncertainty-aware sampling algorithm and the overall active-learning-based diffusion framework. Section~\ref{ER} reports the experimental results, and Section~\ref{Conclusion} concludes this paper.

\section{Problem Formulation and Preliminaries}\label{PF}
\subsection{CGM Reconstruction From Sparse Observations}
If a physical space is discretized into an $H \times W$ grid, the CGM over this physical space can be represented as $X \in \mathbb{R}^{H \times W}$. In practice, obtaining a complete CGM through channel measurements or channel estimation is usually difficult. Therefore, the entire CGM is commonly reconstructed from the physical environment information and sparse observations. In the CGM construction problem, the available physical environment information may include the transmitter location map $L_{\rm TX} \in \mathbb{R}^{H \times W}$, the building map $L_{\rm B} \in \mathbb{R}^{H \times W}$, and the vehicle location map $L_{\rm C} \in \mathbb{R}^{H \times W}$. Specifically, the transmitter location map takes the value of 1 at the transmitter location and 0 elsewhere. The building map takes the value of 1 at building locations and 0 elsewhere. Similarly, the vehicle location map takes the value of 1 at vehicle locations and 0 elsewhere. The sparse observation map $Y \in \mathbb{R}^{H \times W}$ contains the observed values at the observation locations and is set to 0 at unobserved locations. The observation mask $M \in \mathbb{R}^{H \times W}$ takes the value of 1 at the observation locations and 0 at unobserved locations. Therefore, the relationship between the sparse observations and the complete CGM can be expressed as follows:
\begin{equation}
    \label{eq_1}
    Y = M \odot X + N,
\end{equation}
where $\odot$ denotes the Hadamard product and $N$ represents additive white Gaussian noise matrix. Let $f_{\theta}(\cdot)$ denote a learnable reconstruction model parameterized by $\theta$. The predicted CGM $X_{\rm pred}$ is obtained from the sparse observations, the observation mask, and the environmental maps as:
\begin{equation}
    \label{eq_2}
    X_{\rm pred} = f_{\theta}\left(Y,M,L_{\rm TX},L_{\rm B},L_{\rm C}\right).
\end{equation}
To preserve the observed entries while estimating the unobserved ones, the final reconstructed CGM is defined as:
\begin{equation}
    \label{eq_3}
    \hat{X} = M \odot Y + \left(1-M\right)\odot X_{\rm pred}.
\end{equation}

\subsection{Diffusion Model}\label{diffusion model}
Denoising diffusion probabilistic models (DDPMs) \cite{ho2020denoising} formulate the forward noising process and the reverse denoising process as Markov chains. The forward noising process of a diffusion model is fixed, where $\mathbf{x}_{t-1}$ denotes the input data at time step $t-1$, and $\mathbf{x}_{t}$ denotes the noised data at time step $t$. The noising transition from $\mathbf{x}_{t-1}$ to $\mathbf{x}_{t}$ is modeled as follows:
\begin{equation}
    \label{eq_4}
    \mathbf{x}_{t} = \sqrt{1-\beta_t}\mathbf{x}_{t-1} + \sqrt{\beta_t}\boldsymbol{\epsilon}_{t}, \quad \boldsymbol{\epsilon}_{t}\sim\mathcal{N}(\mathbf{0},\mathbf{I}),
\end{equation}
where $\beta_t$ is a predefined noise-scheduling coefficient. Equivalently, the conditional distribution of the forward transition is given by:
\begin{equation}
    \label{eq_5}
    q(\mathbf{x}_{t}\mid\mathbf{x}_{t-1}) = \mathcal{N}\left(\mathbf{x}_{t}; \sqrt{1-\beta_t}\,\mathbf{x}_{t-1},\beta_t\mathbf{I}\right).
\end{equation}
Furthermore, the data at an arbitrary time step can be obtained from the original input data $\mathbf{x}_0$ as follows:
\begin{equation}
    \label{eq_6}
    \mathbf{x}_{t} = \sqrt{\bar{\alpha}_{t}}\,\mathbf{x}_{0} + \sqrt{1-\bar{\alpha}_{t}}\,\bar{\boldsymbol{\epsilon}}_{t},
\end{equation}
where $\bar{\alpha}_{t}=\prod_{s=1}^{t}\alpha_s$, $\alpha_s = 1-\beta_s$. When the number of diffusion steps $\rm T$ is sufficiently large, $\mathbf{x}_{\rm T}$ approximately follows a standard Gaussian distribution.

The core of diffusion models lies in learning the true reverse posterior distribution $q(\mathbf{x}_{t-1}\mid\mathbf{x}_{t})$ from the forward noising process. This true posterior distribution can be simplified as a Gaussian distribution, where the variance is determined by predefined hyperparameters and the only unknown term in the mean is $\mathbf{x}_{0}$. Therefore, a neural network is employed to predict $\mathbf{x}_{0}$, and the loss function is given by:
\begin{equation}
    \label{eq_7}
    \mathcal{L} = \left\|\mathbf{x}_{0} - \mathbf{x}_{\theta}(\mathbf{x}_{t},t)\right\|^{2}.
\end{equation}

The reverse denoising process of a diffusion model corresponds to the generation process. Specifically, an initial sample $\mathbf{x}_{\rm T}$ is first drawn from $\mathcal{N}(\mathbf{0},\mathbf{I})$. Then, $\mathbf{x}_{\rm t-1}$ is sampled from the posterior distribution $q(\mathbf{x}_{t-1}\mid\mathbf{x}_{t})$. After $\rm T$ iterative denoising steps, the final generated result $\mathbf{x}_{0}$ is obtained. The single-step sampling formula according to the posterior distribution is given by:
\begin{equation}
    \label{eq_8}
    \mathbf{x}_{t-1} = \frac{\sqrt{\alpha_t}(1-\bar{\alpha}_{t-1})\mathbf{x}_{t} + \sqrt{\bar{\alpha}_{t-1}}(1-\alpha_t) \mathbf{x}_{\theta}(\mathbf{x}_{t},t)} {1-\bar{\alpha}_{t}}
    + \tilde{\beta}_{t}\boldsymbol{\epsilon}.
\end{equation}

The generation time of DDPM is determined by the number of sampling steps $\rm T$. In practice, $\rm T$ is typically set to 1000, which leads to a long generation time for DDPM. To reduce the CGM construction time and meet the requirement of real-time CGM construction, all sampling algorithms used in this paper are based on the denoising diffusion implicit model (DDIM) \cite{Song2020DenoisingDI}. DDIM relaxes the constraint that the reverse denoising process of DDPM must be Markovian, and thus enables faster sampling while maintaining generation quality. The sampling formula of DDIM is given by:
\begin{equation}
    \label{eq_9}
    \begin{aligned}
    & \tilde{\boldsymbol{\epsilon}}_{\theta} = \frac{\mathbf{x}_{t} - \sqrt{\bar{\alpha}_{t}}\, \mathbf{x}_{\theta}(\mathbf{x}_{t},t)} {\sqrt{1-\bar{\alpha}_{t}}}, \\
    & \mathbf{x}_{t-1} = \sqrt{\bar{\alpha}_{t-1}}\, \mathbf{x}_{\theta}(\mathbf{x}_{t},t) + \sqrt{1-\bar{\alpha}_{t-1}}\, \tilde{\boldsymbol{\epsilon}}_{\theta}.
    \end{aligned}
\end{equation}
Since the above sampling process does not need to follow a Markov chain, the sampled data $\mathbf{x}_{t-1}$ can be obtained from $\mathbf{x}_{t}$ with multiple time steps skipped between them, thereby accelerating the generation process.

Considering the high spatial dimensionality of CGM, 
this paper adopts the latent diffusion model (LDM) \cite{Rombach2021HighResolutionIS} in practice to reduce the training cost. Specifically, in an LDM, the CGM $X$ is first compressed into a latent feature tensor $\mathbf{z}$ by the encoder of a variational autoencoder (VAE). The diffusion model is then trained in the manifold space where the latent feature tensor resides. In the generation stage, the latent feature tensor $\hat{\mathbf{z}}$ is first obtained using the DDIM sampling algorithm, and is then decompressed by the VAE decoder to obtain the predicted CGM $X_{\rm pred}$.

The CGM generation considered in this paper is a condition-guided generation process. The conditions for CGM generation include the transmitter location map, the building map, the vehicle location map, the sparse observation map, and the observation mask. These components are concatenated along the channel dimension to obtain the condition tensor $\mathbf{c}$. In this paper, a cross-attention mechanism \cite{Rombach2021HighResolutionIS} is adopted. Specifically, the condition tensor $\mathbf{c}$ is first encoded by the condition encoder and then injected into the neural network $\mathbf{x}_{\theta}$ through cross-attention.

\subsection{Uncertainty Quantification and Laplace Approximation}\label{LA}
In machine learning, the uncertainty of model predictions generally consists of epistemic uncertainty and aleatoric uncertainty \cite{chan2024estimating}. Aleatoric uncertainty is caused by the inherent noise in the data observation process and cannot be reduced by increasing the number of observations. In contrast, epistemic uncertainty arises from insufficient training data or from the absence of similar samples in the training dataset, i.e., samples that are out of distribution. As a result, the model lacks sufficient knowledge of specific regions during prediction. Epistemic uncertainty represents the uncertainty in model parameters or model functions and can therefore be reduced by acquiring additional observations. Based on the above analysis, epistemic uncertainty should be used to guide sampling. Therefore, it is necessary to quantify the epistemic uncertainty of CGM generated by the diffusion model.

In deep learning, Bayesian neural networks (BNNs) \cite{blundell2015weight} are one of the most commonly used approaches for quantifying epistemic uncertainty. A BNN treats the parameters of the neural network as random variables. When making predictions with a BNN, the predictive distribution can be obtained according to Bayesian inference as follows:
\begin{equation}
    \label{eq_10}
    p(\mathbf{y}^{*}\mid\mathcal{D},\mathbf{x}^{*}) = \int p(\mathbf{y}^{*}\mid\mathbf{x}^{*},\bm{\theta}) p(\bm{\theta}\mid\mathcal{D}) d\bm{\theta},
\end{equation}
where $\bm{\theta}$ denotes the parameters of the Bayesian neural network, $\mathcal{D} = \{(\mathbf{x}_i, \mathbf{y}_i)\}_{i=1}^N$ represents the training dataset, $\mathbf{x}^{*}$ is the new input data outside the training dataset for prediction, and $\mathbf{y}^{*}$ is the corresponding output. The posterior distribution $p(\bm{\theta}\mid\mathcal{D})$ can be obtained according to Bayes' theorem as follows:
\begin{equation}
    \label{eq_11}
    p(\bm{\theta}\mid\mathcal{D}) = \frac{p(\mathcal{D}\mid\bm{\theta})p(\bm{\theta})} {\int p(\mathcal{D}\mid\bm{\theta'})p(\bm{\theta'})d\bm{\theta'}},
\end{equation}
where $p(\mathcal{D}|\bm{\theta}) = \prod_{i=1}^N p(\mathbf{y}_i | \mathbf{x}_i, \bm{\theta})$, the integral term $\int p(\mathcal{D}\mid\bm{\theta'})p(\bm{\theta'})d\bm{\theta'}$ is marginal likelihood, which is generally intractable for neural networks. Consequently, the posterior $p(\bm{\theta}\mid\mathcal{D})$ must be approximated. Common approximation methods include variational inference \cite{hernandez2015probabilistic}, Markov chain Monte Carlo (MCMC) \cite{chen2014stochastic}, deep ensembles \cite{lakshminarayanan2017simple}, MC dropout \cite{gal2016dropout}, and the Laplace approximation \cite{ritter2018scalable}.

Among these methods, the Laplace approximation (LA) does not require model retraining. Instead, it can directly perform uncertainty quantification based on the neural network parameters obtained through conventional point-estimate training. In addition, unlike MCMC, the LA does not need to construct a Markov chain during Bayesian inference, which would otherwise lead to excessive estimation time. Considering that the proposed diffusion model for CGM construction contains a large number of parameters, retraining the model is time-consuming and costly. Moreover, diffusion models usually require a long generation time, and constructing a Markov chain on top of the diffusion sampling process would make the CGM construction time unacceptable. Therefore, in this paper, we employ the LA to perform uncertainty quantification for the diffusion model.

LA approximates the posterior $p(\bm{\theta}\mid\mathcal{D})$ by a Gaussian distribution $\mathcal{N}(\bm{\mu},\bm{\Sigma})$. The key is to determine the mean vector and covariance matrix. Since the marginal likelihood in Eq.\eqref{eq_11} is a constant with respect to $\bm{\theta}$, one may consider the product of the likelihood $p(\mathcal{D}\mid\bm{\theta})$ and the prior $p(\bm{\theta})$:
\begin{equation}
    \label{eq_12}
    \ln p(\bm{\theta})p(\mathcal{D}\mid\bm{\theta}) = \sum_{i=1}^{N} \ln p(\mathbf{y}_{i}\mid\mathbf{x}_{i},\bm{\theta}) + \ln p(\bm{\theta}),
\end{equation}
where the prior $p(\bm{\theta})$ is usually assumed to follow a Gaussian distribution $\mathcal{N}(\mathbf{0},\gamma^{2}\mathbf{I})$ and $\gamma^{2}$ denotes the prior variance. For regression tasks, Eq.\eqref{eq_12} can be written as:
\begin{equation}
    \label{eq_13}
    \ln p(\bm{\theta})p(\mathcal{D}\mid\bm{\theta}) = -\sum_{i=1}^{N} \frac{1}{2\sigma^{2}}\left\|\mathbf{y}_{i} - f_{\bm{\theta}}(\mathbf{x}_{i}) \right\|^{2} - \frac{1}{2}\gamma^{-2}\bm{\theta}^2,
\end{equation}
where $\sigma^{2}$ denotes the noise variance in the regression task, and $f_{\bm{\theta}}(\cdot)$ denotes the prediction function. If we define the loss function $\mathcal{L}(\bm{\theta})$ as follows:
\begin{equation}
    \label{eq_14}
    \mathcal{L}(\bm{\theta}) = \sum_{i=1}^{N} \frac{1}{2\sigma^{2}} \left\| \mathbf{y}_{i} - f_{\bm{\theta}}(\mathbf{x}_{i}) \right\|^{2} + \frac{1}{2}\gamma^{-2}\bm{\theta}^2,
\end{equation}
then,
\begin{equation}
    \label{eq_15}
    p(\bm{\theta})p(\mathcal{D}\mid\bm{\theta}) = \exp\left[-\mathcal{L}(\bm{\theta})\right].
\end{equation}
The loss function in Eq.\eqref{eq_14} consists of an $l_2$ loss term and a regularization term. Thus, training a conventional neural network by minimizing this loss is equivalent to maximizing the posterior probability. The resulting point estimate is the maximum a posteriori (MAP) estimate:
\begin{equation}
    \label{eq_16}
    \bm{\theta}_{\rm MAP} = \arg\max_{\bm{\theta}}\ln p(\bm{\theta}\mid\mathcal{D}).
\end{equation}
Next, a second-order Taylor expansion of $\mathcal{L}(\bm{\theta})$ is performed around $\bm{\theta}_{\rm MAP}$. Since $\mathcal{L}(\bm{\theta})$ is minimized at $\bm{\theta}_{\rm MAP}$, the first-order term $\nabla_{\bm{\theta}}\mathcal{L}(\bm{\theta})$ vanishes. The expansion is
\begin{equation}
    \label{eq_17}
    \mathcal{L}(\bm{\theta}) \approx \mathcal{L}(\bm{\theta}_{\rm MAP}) + \frac{1}{2} (\bm{\theta}-\bm{\theta}_{\rm MAP})^{\mathsf{T}}\left. \nabla_{\bm{\theta}}^{2}\mathcal{L}(\bm{\theta})\right|_{\bm{\theta}_{\rm MAP}} (\bm{\theta}-\bm{\theta}_{\rm MAP}).
\end{equation}
Combining Eq.\eqref{eq_16} and Eq.\eqref{eq_17}, the posterior can be approximated by a Gaussian distribution whose mean and covariance are
\begin{equation}
    \label{eq_18}
    \bm{\mu} = \bm{\theta}_{\rm MAP}, \quad \bm{\Sigma} = \left[\left.\nabla_{\bm{\theta}}^{2}\mathcal{L}(\bm{\theta})\right|_{\bm{\theta}_{\rm MAP}}\right]^{-1}.
\end{equation}
The mean vector is given by the trained neural network parameters, whereas the covariance requires computing the inverse of the Hessian matrix of the loss function. Since neural networks usually contain a large number of parameters, this Hessian matrix is generally difficult to compute directly. Moreover, the Hessian matrix of a general loss function may be indefinite and thus cannot usually serve as a valid covariance matrix. Practical LA therefore relies on simplified curvature approximations, such as the Fisher information matrix or the generalized Gauss-Newton matrix \cite{daxberger2021laplace}. 

However, as the number of network parameters continues to increase, computing the Hessian matrix of the loss function still incurs a prohibitive computational burden, even when simplified approximation methods are adopted. Reference \cite{kristiadi2020being} demonstrated that, when the Laplace approximation is used, the uncertainty quantification results obtained by treating all neural network parameters as random variables are equivalent to those obtained by treating only the last-layer parameters as random variables. Therefore, it is sufficient to compute the approximate posterior distribution of only the last-layer parameters, which further reduces the computational complexity. This approach is also referred to as the last-layer Laplace approximation (LLLA). In this paper, LLLA is also adopted to quantify the uncertainty of the results generated by the diffusion model.

\section{Epistemic Uncertainty Quantification Based on Diffusion Model}\label{EUQBDM}
\subsection{Single Reverse Diffusion Uncertainty Quantification}\label{SRUQ}
\begin{figure}[!htbp]
\centering
\includegraphics[width=2.8in]{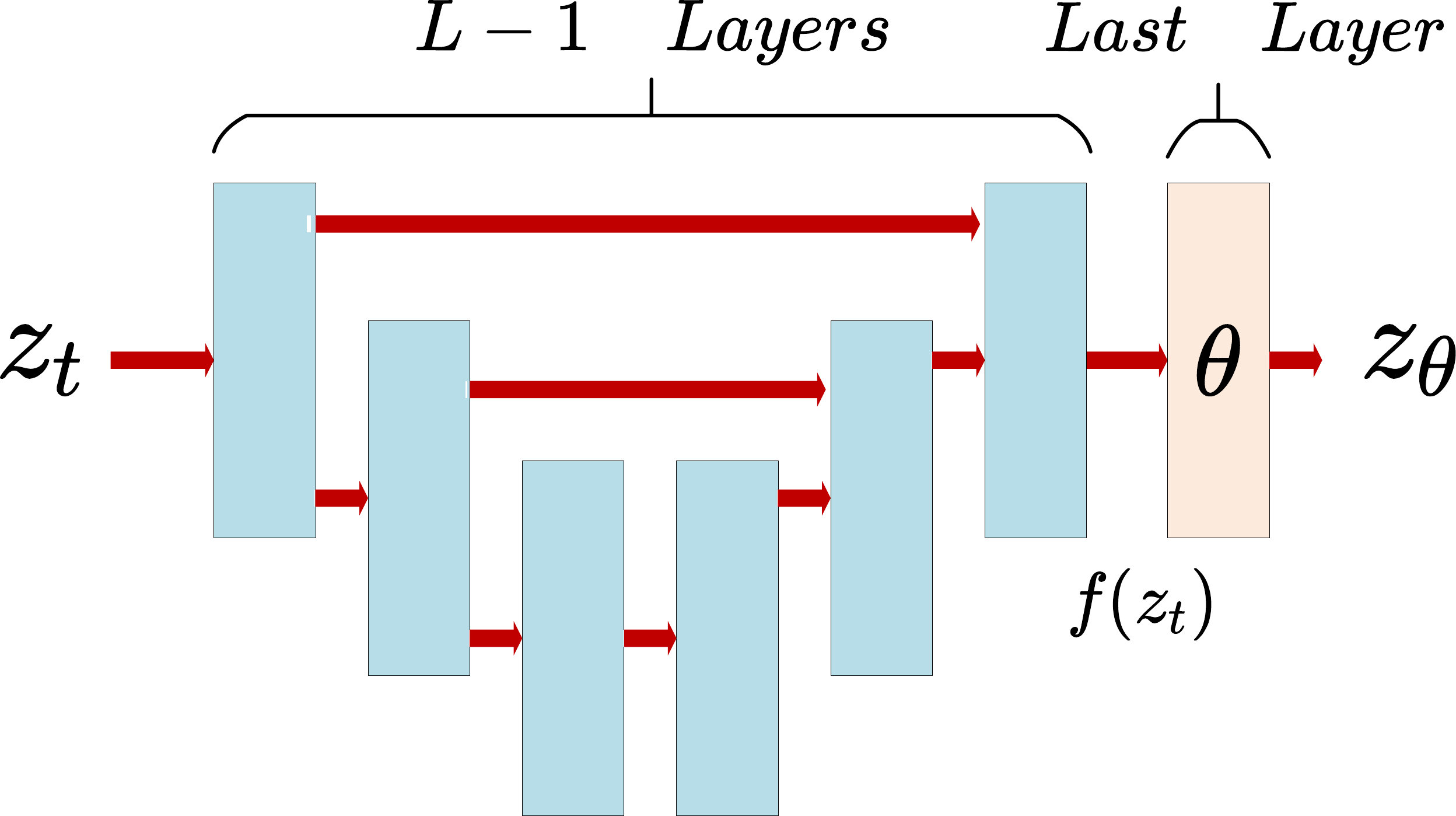}
\caption{Schematic illustration of the U-Net architecture used in the diffusion model.}
\label{fig_1}
\end{figure}

The proposed diffusion-based active learning framework for CGM construction adopts the latent diffusion model described in Section~\ref{diffusion model} and the DDIM sampling algorithm. In addition, the posterior distribution of the model parameters is estimated using the LLLA introduced in Section~\ref{LA}. The U-Net architecture of the latent diffusion model is illustrated in Fig. \ref{fig_1}.

Suppose that at a given time step, the input latent feature tensor $\mathbf{z}_{t}$ has dimensions of $\mathbb{R}^{\rm c \times hw}$ by vectorizing the spatial dimensions. Let the utilized U-Net architecture consist of $L$ layers, and denote the mapping function of the first $L-1$ layers as $f(\cdot)$. The output of the first $L-1$ layers, $f(\mathbf{z}_{t})$, has dimensions of $\mathbb{R}^{\rm c_{\dim} \times hw}$, where $\rm c_{\dim}$ is the feature dimension before the final layer, with $\rm c_{\dim}>c$. Following the mapping by the final convolutional layer, the ultimate output dimension of the U-Net $\mathbf{z}_{\theta}$ is $\mathbb{R}^{\rm c \times hw}$. Since the kernel size of the last convolutional layer is one and no bias term is used, it has the same functional form as a fully connected linear layer. Consequently, the weight parameters of the final U-Net layer are of dimension $\rm c \times \rm c_{\dim}$, and their posterior probability distribution, approximated via the LLLA, is formulated as:
\begin{equation}
    \label{eq_19}
    p(\bm{\theta}\mid\mathcal{D})\overset{\mathrm{L.A.}}{\simeq}\mathcal{N}\left(\bm{\theta};\bm{\theta}_{\rm MAP},\Sigma_{\theta}\right).
\end{equation}
Accordingly, the relation between the U-Net output $\mathbf{z}_{\theta}$ and the input $\mathbf{z}_{t}$ can be expressed as
\begin{equation}
    \label{eq_20}
    \mathbf{z}_{\theta} = \bm{\theta} f(\mathbf{z}_{t}).
\end{equation}
According to Eq.\eqref{eq_9} in Section~\ref{diffusion model}, the single-step sampling formula of DDIM is given by:
\begin{equation}
    \label{eq_21}
    \mathbf{z}_{t-1} = A_t \mathbf{z}_{t}  + B_t \mathbf{z}_{\theta}(\mathbf{z}_{t},t),
\end{equation}
where $A_t = \frac{\sqrt{1-\bar{\alpha}_{t-1}}}{\sqrt{1-\bar{\alpha}_{t}}}$ and $B_t = \sqrt{\bar{\alpha}_{t-1}} - \frac{\sqrt{1-\bar{\alpha}_{t-1}}\sqrt{\bar{\alpha}_{t}}}{\sqrt{1-\bar{\alpha}_{t}}}$. Combining \eqref{eq_20} and \eqref{eq_21} yields
\begin{equation}
    \label{eq_22}
    \mathbf{z}_{t-1} = A_t\mathbf{z}_{t} + B_t \bm{\theta} f(\mathbf{z}_{t}).
\end{equation}
By expressing the relationship in Eq.\eqref{eq_22} in a simplified form as $\mathbf{z}_{t-1}=g(\bm{\theta})$, it is evident that $g(\cdot)$ is a non-linear function.

Performing a first-order Taylor expansion of $g(\bm{\theta})$ around $\bm{\theta}_{\rm MAP}$ yields:
\begin{equation}
    \label{eq_23}
    \mathbf{z}_{t-1} \approx g(\bm{\theta}_{\rm MAP}) + \left.\frac{d g(\bm{\theta})}{d\bm{\theta}}\right|_{\bm{\theta}_{\rm MAP}}\left(\bm{\theta}-\bm{\theta}_{\rm MAP}\right).
\end{equation}
Let $\bm{J}_{t-1} = \left.\frac{d g(\bm{\theta})}{d\bm{\theta}}\right|_{\bm{\theta}_{\rm MAP}}$, the expectation $\mathbb{E}_{p(\bm{\theta}\mid\mathcal{D})}[\mathbf{z}_{t-1}]$ is
\begin{equation}
    \label{eq_24}
    \begin{aligned}
    &\mathbb{E}_{p(\bm{\theta}\mid\mathcal{D})}[\mathbf{z}_{t-1}] \approx \mathbb{E}_{p(\bm{\theta}\mid\mathcal{D})}
    \left[g(\bm{\theta}_{\rm MAP}) + \bm{J}_{t-1} (\bm{\theta}-\bm{\theta}_{\rm MAP})\right] \\
    & = g(\bm{\theta}_{\rm MAP}) + \bm{J}_{t-1}\left(\mathbb{E}_{p(\bm{\theta}\mid\mathcal{D})}[\bm{\theta}] - \bm{\theta}_{\rm MAP}\right) = g(\bm{\theta}_{\rm MAP}).
    \end{aligned}
\end{equation}
Similarly, the covariance $\operatorname{Cov}_{p(\bm{\theta}\mid\mathcal{D})}[\mathbf{z}_{t-1}]$ is
\begin{equation}
    \label{eq_25}
    \begin{aligned}
    \operatorname{Cov}_{p(\bm{\theta}\mid\mathcal{D})}[\mathbf{z}_{t-1}] &= \mathbb{E}_{p(\bm{\theta}\mid\mathcal{D})}\left[\left(\mathbf{z}_{t-1}-\mathbb{E}_{p(\bm{\theta}\mid\mathcal{D})}[\mathbf{z}_{t-1}]\right)\left(\mathbf{z}_{t-1}-\mathbb{E}_{p(\bm{\theta}\mid\mathcal{D})}[\mathbf{z}_{t-1}]\right)^{\top}\right] \\
    &\approx \mathbb{E}_{p(\bm{\theta}\mid\mathcal{D})}\left[(\bm{J}_{t-1}(\bm{\theta}-\bm{\theta}_{\rm MAP}))(\bm{J}_{t-1}(\bm{\theta}-\bm{\theta}_{\rm MAP}))^{\top}\right] \\
    & = \bm{J}_{t-1}\mathbb{E}_{p(\bm{\theta}\mid\mathcal{D})}\left[(\bm{\theta}-\bm{\theta}_{\rm MAP})(\bm{\theta}-\bm{\theta}_{\rm MAP})^{\top}\right] \bm{J}_{t-1}^{\top}.
    \end{aligned}
\end{equation}
Because $\boldsymbol{\Sigma}_{\theta} = \mathbb{E}_{p(\bm{\theta}\mid\mathcal{D})}\left[(\bm{\theta}-\bm{\theta}_{\rm MAP})(\bm{\theta}-\bm{\theta}_{\rm MAP})^{\top}\right]$, we can get
\begin{equation}
    \label{eq_26}
    \operatorname{Cov}_{p(\bm{\theta}\mid\mathcal{D})}[\mathbf{z}_{t-1}] = \bm{J}_{t-1} \bm{\Sigma}_{\theta} \bm{J}_{t-1}^{\top}.
\end{equation}
Therefore, $\mathbf{z}_{t-1}$ can be approximated by a Gaussian distribution $\mathcal{N}(\bm{\mu}_{t-1}, \bm{\Sigma}_{t-1})$, where the mean vector and the covariance matrix are
\begin{equation}
    \label{eq_27}
    \begin{aligned}
    &\bm{\mu}_{t-1}= A_t\bm{\mu}_{t} + B_t\bm{\theta}_{\rm MAP}f(\bm{\mu}_{t}) \\
    &\bm{\Sigma}_{t-1}=\bm{J}_{t-1}\bm{\Sigma}_{\theta}\bm{J}_{t-1}^{\top}.
    \end{aligned}
\end{equation}

Following the same procedure as the DDIM sampling algorithm, the approximate Gaussian distribution of $\mathbf{z}_{0}$ can be obtained by iterating for $\rm T$ steps, thereby yielding the epistemic uncertainty quantification via a single reverse diffusion process. The core to realizing this iterative computation is the evaluation of the Jacobian matrix $\bm{J}_{t-1}$ in Eq.\eqref{eq_26}.

According to Eq.\eqref{eq_21}, $\bm{J}_{t-1}$ can be obtained as follows:
\begin{equation}
    \label{eq_28}
    \begin{aligned}
    &\bm{J}_{t-1} = \left. \frac{d \mathbf{z}_{t-1}}{d \bm{\theta}} \right|_{\bm{\theta}_{\rm MAP}} = A_t \left. \frac{d \mathbf{z}_t}{d \bm{\theta}} \right|_{\bm{\theta}_{\rm MAP}} + B_t \left. \left( (\mathbf{I}_{hw} \otimes \bm{\theta}) \frac{\partial \mathrm{vec}(f(\mathbf{z}_t))}{\partial \mathrm{vec}(\bm{\theta})} + f(\mathbf{z}_t)^{\top} \otimes \mathbf{I}_c \right) \right|_{\bm{\theta}_{\rm MAP}}.
    \end{aligned}
\end{equation}
Subsequently, based on the chain rule of differentiation, it can be deduced that:
\begin{equation}
    \label{eq_29}
    \begin{aligned}
    \bm{J}_{t-1} &= A_t \bm{J}_t + B_t\left. \left( (\mathbf{I}_{hw} \otimes \bm{\theta}) \frac{\partial \mathrm{vec}(f(\mathbf{z}_t))}{\partial \mathrm{vec}(\mathbf{z}_t)}\frac{\partial \mathrm{vec}(\mathbf{z}_t)}{\partial \mathrm{vec}(\bm{\theta})} + f(\mathbf{z}_t)^{\top} \otimes \mathbf{I}_c \right) \right|_{\bm{\theta}_{\rm MAP}} \\
    &= A_t\bm{J}_t + B_t (\mathbf{I}_{hw} \otimes \bm{\theta}_{\rm MAP}) \frac{\partial \mathrm{vec}(f(\mathbf{z}_t))}{\partial \mathrm{vec}(\mathbf{z}_t)} \bm{J}_t \\
    &+ B_t f(\mathbf{z}_t)^{\top} \otimes \mathbf{I}_c.
    \end{aligned}
\end{equation}
In Eq.\eqref{eq_29}, the matrices involved have extremely large dimensions.
Therefore, explicitly computing Eq.\eqref{eq_29} incurs substantial computational time and high GPU memory consumption. To mitigate these issues, two implementation techniques are adopted.

The first issue is the computation of $\frac{\partial \mathrm{vec}(f(\mathbf{z}_t))}{\partial \mathrm{vec}(\mathbf{z}_t)} \bm{J}_t$. Computing this expression explicitly would require not only constructing a massive Jacobian matrix but also performing the multiplication of two large-scale matrices. However, by examining this expression, we observe that it corresponds to a typical directional-derivative projection problem, which can be computed using the Jacobian-vector product (JVP). Based on the automatic differentiation mechanism in PyTorch, the JVP can be obtained through a single forward-mode evaluation of the neural network $f(\cdot)$. Therefore, this term can be computed by performing a batched JVP only once, without explicitly constructing the full Jacobian matrix and then multiplying it with the subsequent matrix.


Second, explicitly computing the Kronecker product of $\mathbf{I}_{hw}$ and $\bm{\theta}_{\rm MAP}$ before multiplying it by the subsequent matrix would still necessitate handling prohibitively large matrix computations. Here, by exploiting the diagonal structure of $\mathbf{I}_{hw}$, we use the Einstein summation convention to obtain the final result, thereby significantly reducing the dimensionality of tensor computations. 

Incorporating the two aforementioned techniques allows us to efficiently execute the $\rm T$-step iterative computation via Eq.\eqref{eq_27}, thereby obtaining $\bm{\mu}_0$ and $\bm{J}_0$. Since this work adopts a LDM, the derived distribution is an approximate Gaussian distribution of the latent feature tensor. Consequently, it is still necessary to employ Monte Carlo sampling to compute the variance of the final generated output after passing through the decoder. Sampling $\mathbf{z}_{0,i}$ directly from $\mathcal{N}(\bm{\mu}_{0},\bm{\Sigma}_{0})$ would require computing $\bm{\Sigma}_{0}$ and performing Cholesky decomposition on it. This is costly because forming $\bm{\Sigma}_{0}$ involves large-scale matrix multiplications. Moreover, $\bm{\Sigma}_{0}$ may not be strictly positive definite. To avoid this issue, Cholesky decomposition is first applied to the positive definite matrix $\bm{\Sigma}_{\theta}$:
\begin{equation}
    \label{eq_31}
    \bm{\Sigma}_{\theta}=\bm{S}\bm{S}^{\top},
\end{equation}
where $\bm{S}$ is a lower triangular matrix. Then,
\begin{equation}
    \label{eq_32}
    \bm{\Sigma}_{0}=\bm{J}_{0}\bm{\Sigma}_{\theta}\bm{J}_{0}^{\top}=\bm{J}_{0}\bm{S}\bm{S}^{\mathsf{T}}\bm{J}_{0}^{\mathsf{T}}.
\end{equation}
Therefore, the sample $\mathbf{z}_{0,i}$ can be drawn using the following equation:
\begin{equation}
    \label{eq_33}
    \mathbf{z}_{0,i} = \bm{\mu}_{0}+\bm{J}_{0}\bm{S}\bm{\epsilon}_{i},\quad\bm{\epsilon}_{i}\sim\mathcal{N}(\mathbf{0},\mathbf{I}_{chw}).
\end{equation}
This strategy not only effectively mitigates the dimensionality of the matrix multiplications but also guarantees that the drawn samples $\mathbf{z}_{0,i}$ strictly follow the Gaussian distribution $\mathcal{N}(\bm{\mu}_{0},\bm{\Sigma}_{0})$. Finally, if Monte Carlo sampling is employed to generate $M$ samples, which are subsequently passed through the decoder to yield $M$ corresponding outputs $\mathbf{x}_{0,i}$, the variance of $\mathbf{x}_0$ can be formulated as:
\begin{equation}
    \label{eq_34}
    \operatorname{Var}(\mathbf{x}_{0}) = \frac{1}{M}\sum_{i=1}^{M}\left(\mathbf{x}_{0,i}-\bar{\mathbf{x}}_{0}\right)^{2},\quad\bar{\mathbf{x}}_{0}=\frac{1}{M}\sum_{i=1}^{M}\mathbf{x}_{0,i}.
\end{equation}

\begin{algorithm}
\caption{Single Reverse Diffusion Uncertainty Quantification}\label{alg1}
\begin{algorithmic}[1]
\Require Trained $L-1$ neural network $f(\cdot)$, Trained last-layer parameters $\bm{\theta}_{\rm MAP}$, Decoder $\rm D$, DDIM timesteps $\rm T$, Monte Carlo sample size $M$, 
\State Get $p(\bm\theta\mid \mathcal{D})=\mathcal{N}\left(\bm\theta ; \bm{\theta}_{\rm MAP}, \bm{\Sigma}_\theta\right)$ via LLLA
\State $\bm{S} = \operatorname{Cholesky}(\bm{\Sigma}_\theta)$
\State Let $A_t = \frac{\sqrt{1 - \bar{\alpha}_{t-1}}}{\sqrt{1 - \bar{\alpha}_t}}$, $B_t = \sqrt{\bar{\alpha}_{t-1}} - \frac{\sqrt{1 - \bar{\alpha}_{t-1}} \sqrt{\bar{\alpha}_t}}{\sqrt{1 - \bar{\alpha}_t}}$
\State $\mathbf{z}_{\rm T} \sim \mathcal{N}(\bm{0}, \mathbf{I})$
\State Initialize $\bm{\mu}_{\rm T-1} = A_{\rm T} \mathbf{z}_{\rm T} + B_{\rm T} \bm{\theta}_{\rm MAP} f(\mathbf{z}_{\rm T})$
\State Initialize $\bm{J}_{\rm T-1} = B_{\rm T}f(\mathbf{z}_{\rm T})^\top \otimes \mathbf{I}_c$
\For{$t = {\rm T-1} \to 1$ }
    \State $\bm{\mu}_{\rm t-1} = A_{\rm t} \bm{\mu}_{\rm t} + B_{\rm t} \bm{\theta}_{\text{MAP}} f(\bm{\mu}_{\rm t})$
    \State
    $\begin{aligned}
    &\bm{J}_{\rm t-1} = A_{\rm t} \bm{J}_{\rm t} + B_{\rm t} \left( \mathbf{I}_{hw} \otimes \bm{\theta}_{\text{MAP}} \right) \frac{\partial \operatorname{vec}\left( f\left( \bm{\mu}_{\rm t} \right) \right)}{\partial \operatorname{vec}(\bm{\mu}_{\rm t})} \bm{J}_{\rm t} \\
    &+ B_{\rm t} f\left( \bm{\mu}_{\rm t} \right)^T \otimes \mathbf{I}_c
    \end{aligned}$
\EndFor
\For{$i = 1 \to M$ }
    \State ${\mathbf{z}_{0,i}}={\bm{\mu}_{0}}+{\bm{J}_{0}}\bm{S}\bm{\varepsilon}, \bm{\varepsilon} \sim \mathcal{N}\left(0,\mathbf{I}_{chw} \right)$
    \State $\mathbf{x}_{0,i}=\rm D(\mathbf{z}_{0,i})$
\EndFor
\State $\operatorname{Var}(\mathbf{x}_0)=\frac{1}{M}{\sum\limits_{i=1}^{M}{{{\left(\mathbf{x}_{0,i}-{{{\bar{\mathbf{x}}}}_{0}} \right)}^{2}}}}$, ${{\bar{\mathbf{x}}}_{0}}=\frac{1}{M}\sum\limits_{i=1}^{M}{{\mathbf{x}}_{0,i}}$
\State \Return $\operatorname{Var}(\mathbf{x}_0)$
\end{algorithmic}
\end{algorithm}

The complete algorithmic procedure is summarized in Algorithm \ref{alg1}. Since this method computes the epistemic uncertainty of CGM through a single reverse diffusion process, it is referred to as the single reverse diffusion uncertainty quantification method.

\subsection{Multiple Reverse Diffusion Uncertainty Quantification}\label{MRQU}
The method described in Section~\ref{SRUQ} requires an additional computation of the matrix $\bm{J}_{\rm t-1}$ at each time step of the reverse diffusion process. 
The computational time increases rapidly with the number of DDIM sampling steps $\rm T$. To reduce the time required to obtain the CGM prediction variance map while avoiding excessive computational resource consumption, we further propose a multiple reverse diffusion uncertainty quantification method.

Since the DDIM sampling algorithm is a deterministic sampling process, for a well-trained diffusion model with parameters $\bm{\theta}_{\rm MAP}$, the entire sampling process can be represented by a function $g_{\bm{\theta}}(\cdot)$. Therefore, the generation of a new sample can be expressed as follows:
\begin{equation}
    \label{eq_35}
    \mathbf{x}_{0} = g_{\bm{\theta}}\left(\mathbf{x}_{\rm T},\mathbf{c}\right),
\end{equation}
where $\mathbf{x}_{\rm T}$ denotes the fixed initial noise sampled from $\mathcal{N}(\bm{0},\mathbf{I})$ and $\mathbf{c}$ denotes the guidance condition of the diffusion model. The probability distribution $p(\mathbf{x}_{0}\mid\mathbf{x}_{\rm T},\mathbf{c},\bm{\theta})$ is therefore a Dirac distribution, which can be expressed as:
\begin{equation}
    \label{eq_36}
    p(\mathbf{x}_{0}\mid\mathbf{x}_{\rm T},\mathbf{c},\bm{\theta})=p(\mathbf{x}_{0}\mid g_{\bm{\theta}}(\mathbf{x}_{\rm T},\mathbf{c}))
    =\delta\left(\mathbf{x}_{0}-g_{\bm{\theta}}(\mathbf{x}_{\rm T},\mathbf{c})\right).
\end{equation}

For a diffusion model trained on the dataset $\mathcal{D}$, we aim to compute the epistemic uncertainty of the generated result for a new input, which corresponds to a new guidance condition $\mathbf{c}^*$ in the conditional diffusion model. This uncertainty is characterized by the variance of the probability distribution $p(\mathbf{x}_{0}\mid\mathbf{x}_{\rm T},\mathbf{c}^{*},\bm{\theta})$. According to the Bayesian predictive inference formulation, the probability distribution $p(\mathbf{x}_{0}\mid\mathbf{x}_{\rm T},\mathbf{c}^{*},\bm{\theta})$ can be written as follows:
\begin{equation}
    \label{eq_37}
    \begin{aligned}
    p(\mathbf{x}_{0}\mid\mathbf{x}_{T},\mathbf{c}^{*},\mathcal{D}) &=\int p(\mathbf{x}_{0},\bm{\theta}\mid \mathbf{x}_{T},\mathbf{c}^{*},\mathcal{D})d\bm{\theta}\\
    &=\int p(\mathbf{x}_{0} \mid \bm{\theta},\mathbf{x}_{T},\mathbf{c}^{*},\mathcal{D}) p(\bm{\theta} \mid \mathbf{x}_{T},\mathbf{c}^{*},\mathcal{D})d\bm{\theta}.
    \end{aligned}
\end{equation}
Since $p(\mathbf{x}_{0}\mid\bm{\theta},\mathbf{x}_{\rm T},\mathbf{c}^*,\mathcal{D})=p(\mathbf{x}_{0}\mid\bm{\theta},\mathbf{x}_{\rm T},\mathbf{c}^*)$ and $p(\bm{\theta}\mid\mathbf{x}_{T},\mathbf{c}^*,\mathcal{D})=p(\bm{\theta}\mid\mathcal{D})$, Eq.\eqref{eq_37} can be rewritten as:
\begin{equation}
    \label{eq_38}
    \begin{aligned}
    p(\mathbf{x}_{0}\mid\mathbf{x}_{\rm T},\mathbf{c}^*,\mathcal{D})&=\int p(\mathbf{x}_{0}\mid\bm{\theta},\mathbf{x}_{\rm T},\mathbf{c}^*)p(\bm{\theta}\mid\mathcal{D})d\bm{\theta}\\
    &=\mathbb{E}_{p(\bm{\theta}\mid\mathcal{D})}\left[p(\mathbf{x}_{0}\mid g_{\bm{\theta}}(\mathbf{x}_{\rm T},\mathbf{c}^*))\right].
    \end{aligned}
\end{equation}
Since the integral in the above equation lacks a closed-form solution, it must be evaluated using Monte Carlo methods. First, we apply LLLA to obtain a Gaussian approximation for the posterior distribution of the neural network's final layer parameters. Subsequently, $M$ sets of parameters are sampled from this approximate posterior distribution. The Monte Carlo approximation for the expectation in Eq.\eqref{eq_38} can be expressed as:
\begin{equation}
    \label{eq_39}
    \begin{aligned}
    \mathbb{E}_{p(\bm{\theta}\mid\mathcal{D})}\left[p(\mathbf{x}_{0}\mid g_{\bm{\theta}}(\mathbf{x}_{\rm T},\mathbf{c}^*))\right]
    =\frac{1}{M}\sum_{i=1}^{M}\delta\left(\mathbf{x}_{0}-g_{\bm{\theta}_{i}}(\mathbf{x}_{\rm T},\mathbf{c}^{*})\right),
    \end{aligned}
\end{equation}
where $g_{\bm{\theta}_{i}}(\cdot)$ denotes the DDIM sampling function when the $i$-th sampled parameter set is loaded into the final layer of the neural network. The expectation in Eq.\eqref{eq_39} can be expressed as a mixture of $M$ Dirac delta distributions. Since the variance within each Dirac distribution is zero, the total uncertainty stems entirely from the output variations caused by different parameter samples. Consequently, by employing the moment matching method, Eq.\eqref{eq_39} can be approximated as:
\begin{equation}
    \label{eq_40}
    \begin{aligned}
    \mathbb{E}_{p(\bm{\theta}\mid\mathcal{D})}\left[p(\mathbf{x}_{0}\mid g_{\bm{\theta}}(\mathbf{x}_{\rm T},\mathbf{c}^{*}))\right] 
    \approx \mathcal{N}\left(\bar{\mathbf{g}},\frac{1}{M}\sum_{i=1}^{M}\left(g_{\bm{\theta}_{i}}(\mathbf{x}_{\rm T},\mathbf{c}^{*})^2-\bar{\mathbf{g}}^2\right)\right),
    \end{aligned}
\end{equation}
where
\begin{equation}
    \label{eq_41}
    \bar{\mathbf{g}} = \frac{1}{M} \sum_{i=1}^{M} g_{\bm{\theta}_{i}}(\mathbf{x}_{\rm T},\mathbf{c}^{*})
\end{equation}
and the variance of $\mathbf{x}_0$ can be formulated as:
\begin{equation}
    \label{eq_42}
    \operatorname{Var}(\mathbf{x}_{0}) = \operatorname{diag} \left\{\frac{1}{M}\sum_{i=1}^{M}\left(g_{\bm{\theta}_{i}}(\mathbf{x}_{\rm T},\mathbf{c}^{*})^2-\bar{\mathbf{g}}^2\right)\right\}.
\end{equation}

\begin{algorithm}
\caption{Multiple Reverse Diffusion Uncertainty Quantification}\label{alg2}
\begin{algorithmic}[1]
\Require Monte Carlo sample size $M$, Number of last-layer parameters loaded in parallel $P$, Trained neural network $\mathbf{z}_\theta$, condition prompt $c^{\ast}$
\State  Get $p(\bm{\theta}\mid \mathcal{D})=\mathcal{N}\left(\bm{\theta}; \bm{\theta}_{\rm MAP}, \bm{\Sigma}_\theta\right)$ via LLLA
\State $\mathbf{z}_{\rm T} \sim \mathcal{N}(\mathbf{0}, \mathbf{I})$
\For{$m = 1 \to M/P$ }
    \State Sample $P$ values of $\bm{\theta}_i$ from $p(\bm{\theta} \mid \mathcal{D})$
    \State Load the parameters $\bm{\theta}_i$ into the network $\mathbf{z}_\theta$, obtain 
    \Statex \hspace{\algorithmicindent}
    the corresponding DDIM sampling function $g_{\bm{\theta}_i}(\cdot)$
    \State Generate $P$ samples through a single DDIM reverse
    \Statex \hspace{\algorithmicindent}
    diffusion process: $\mathbf{x}_{0,i}=g_{\bm{\theta}_i}(\mathbf{z}_{\rm T}, c^{\ast})(i=1,2,\ldots,P)$
\EndFor
\State $\bar{\mathbf{x}}_0=\frac{1}{M} \sum_{i=1}^{M} \mathbf{x}_{0,i}$
\State $\operatorname{Var}(\mathbf{x}_0)=\frac{1}{M} \sum_{i=1}^{M} \mathbf{x}_{0,i}^2-\bar{\mathbf{x}}_0^2$
\State \Return $\operatorname{Var}(\mathbf{x}_0)$
\end{algorithmic}
\end{algorithm}

The above method requires multiple reverse diffusion processes; therefore, we refer to it as the multiple reverse diffusion uncertainty quantification method. Although the above method requires sampling multiple generated samples $\mathbf{x}_{0,i}$, it can be computationally faster than the single reverse diffusion uncertainty quantification method. This is because, first, this method does not require computing the complicated matrix $\bm{J}_{t-1}$ at each denoising step during the reverse sampling process; therefore, each sampling trajectory can be executed more efficiently.
Second, the Monte Carlo method is inherently parallelizable and can be accelerated through parallel computation. Since the LLLA is adopted in this paper, the models corresponding to different sampled functions $g_{\bm{\theta}_{i}}(\cdot)$ differ only in their last-layer parameters, while the remaining network parameters are shared. Therefore, parallel implementation does not require simultaneously loading multiple complete models; instead, only the sampled last-layer parameters need to be loaded. As a result, the proposed method does not introduce substantial additional computational overhead in parallel implementation and naturally benefits from parallel acceleration.

\begin{figure}[!htbp]
\centering
\includegraphics[width=2.8in]{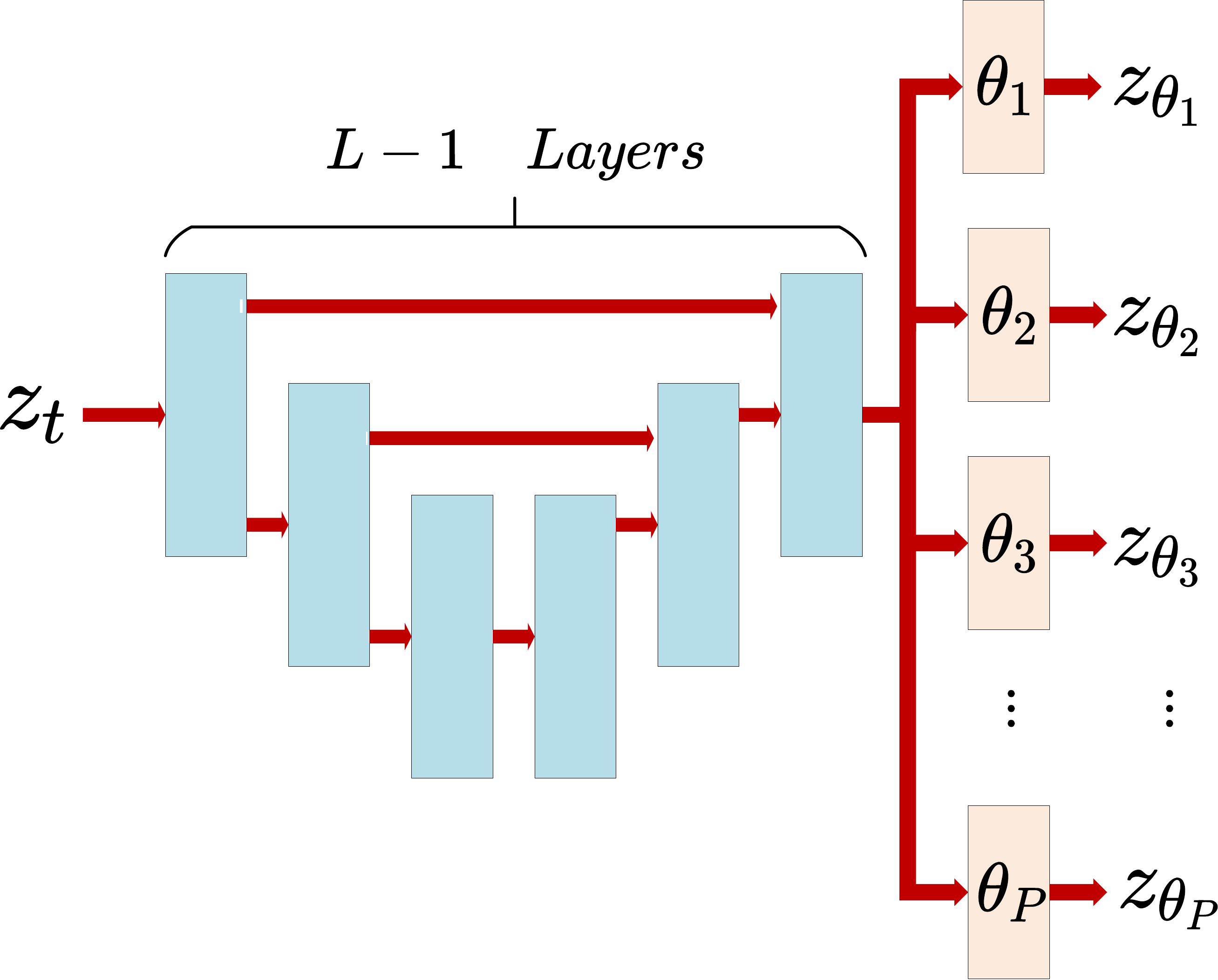}
\caption{Schematic illustration of parallel loading of model parameters.}
\label{fig_2}
\end{figure}

In the implementation, to balance GPU memory consumption and computational time, we adopt a partially parallelized computation strategy, as illustrated in Fig. \ref{fig_2}. Specifically, $P$ sets of last-layer network parameters are loaded simultaneously to compute the sampled outputs in parallel. In this way, $P$ samples $\mathbf{x}_{0,i}(i=1,2,\ldots,P)$ can be obtained within a single reverse diffusion process, and the Monte Carlo sampling can be completed after $M/P$ reverse diffusion processes. The complete procedure of the multiple reverse diffusion uncertainty quantification algorithm is summarized in Algorithm \ref{alg2}.

\section{Uncertainty-Aware Sampling Algorithm and Active-Learning-Based Diffusion Framework for CGM Construction}\label{UASA}
\subsection{Uncertainty-Aware Sampling Algorithm}
\begin{figure}[!htbp]
\centering
\subfloat[]{
    \centering
    \includegraphics[width=0.9\columnwidth]{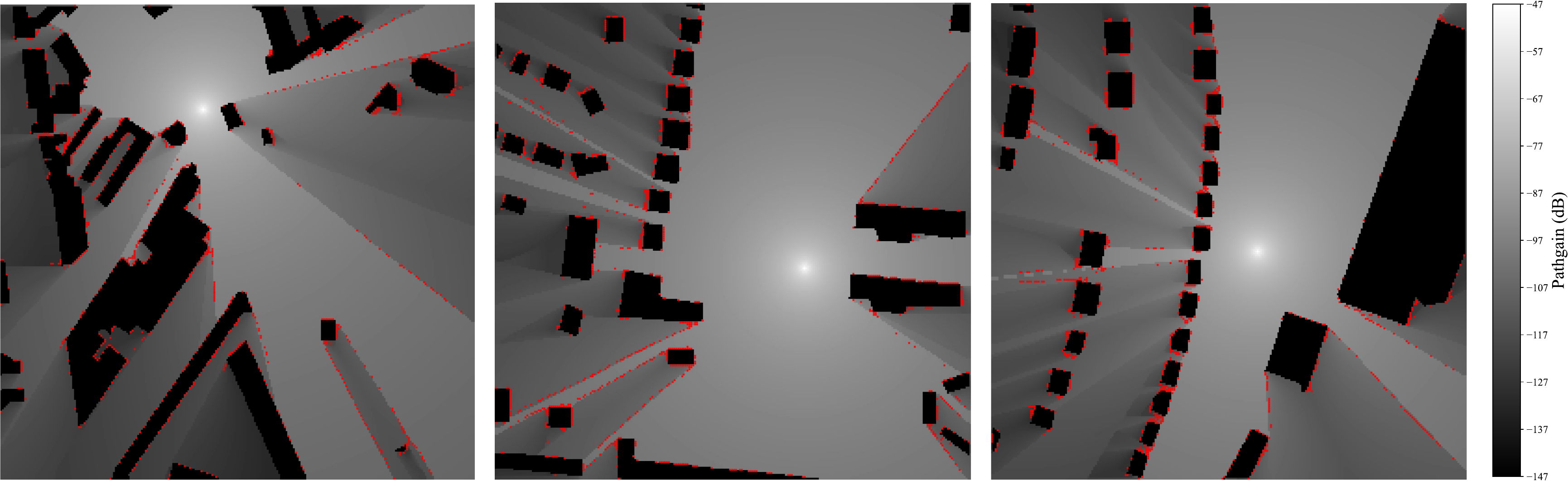}
    \label{fig_3a}
}
\hfill
\subfloat[]{
    \centering
    \includegraphics[width=0.9\columnwidth]{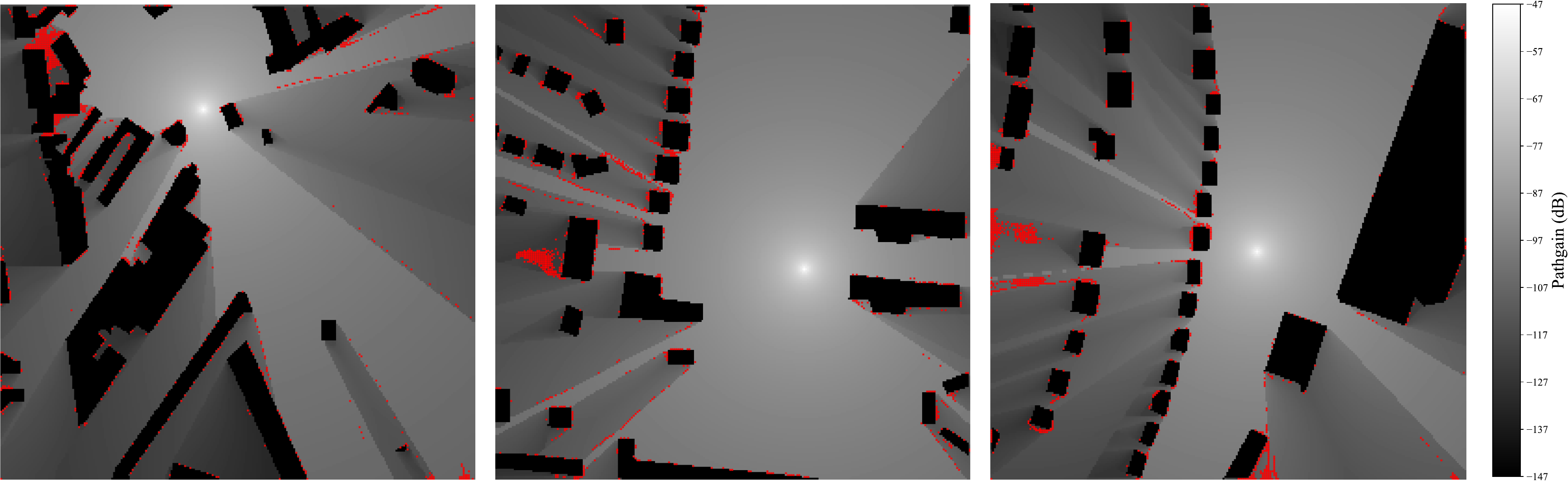}
    \label{fig_3b}
}
\caption{Spatial distribution of sampling points selected by top-K variance. (a) CGM prediction variance map obtained by Algorithm 1. (b) CGM prediction variance map obtained by Algorithm 2.}
\label{fig_3}
\end{figure}

After obtaining the CGM prediction variance map, directly selecting the $K$ locations with the largest variances may lead to overly clustered sampling points, resulting in insufficient spatial coverage and excessive information redundancy. This phenomenon is illustrated in Fig. \ref{fig_3a} and Fig. \ref{fig_3b}, where the red markers denote the selected sampling points. 
It can also be observed that the top-K sampling points selected from the variance map obtained by Algorithm \ref{alg2} exhibit a higher degree of spatial concentration than those selected from the variance map obtained by Algorithm \ref{alg1}.

To ensure that the selected sampling points are not excessively clustered while still favoring locations with large prediction variance, we design the measurement location selection algorithm based on the following principles. First, locations with larger variance should have a higher probability of being selected. Second, a minimum distance should be maintained between any two selected sampling points to avoid insufficient spatial coverage. The proposed measurement location selection method is designed as follows. After obtaining the CGM prediction variance map $\rm V$ based on the initial sampling points set $\rm S_0$ using the algorithms described in Section~\ref{SRUQ} or Section~\ref{MRQU}, we first normalize it to obtain $\rm \tilde{V}$. Then, the top $q\%$ locations with the largest variances in $\rm \tilde{V}$ are selected to form the candidate pool $\rm C$. Each candidate location is then sampled according to a probability distribution $\rm P$. Specifically, if the candidate pool $\rm C$ contains $n$ candidate locations, the probability that the i-th candidate point $p_i(i=1,2,\ldots,n)$ is selected can be expressed as:

\begin{equation}
    \label{eq_43}
    {\rm P}(p_i) = \frac{\rm \tilde{V}_i}{\sum_{j=1}^{n} \rm \tilde{V}_j}.
\end{equation}

\begin{figure}[!htbp]
\centering
\subfloat[]{
    \centering
    \includegraphics[width=0.9\columnwidth]{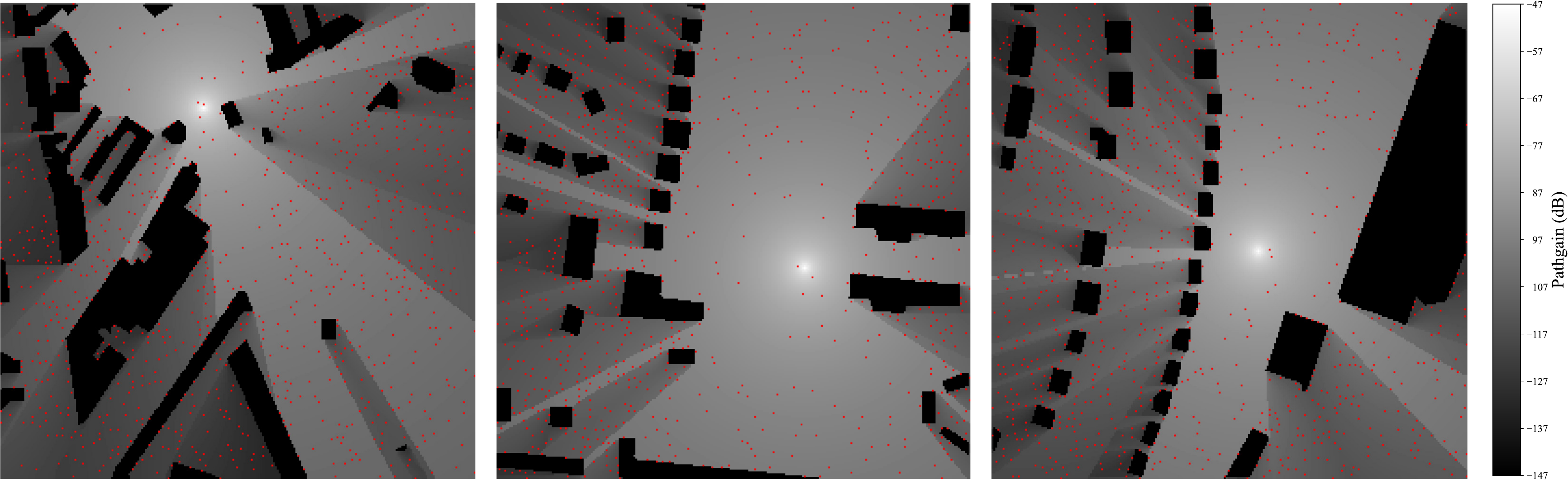}
    \label{fig_4a}
}
\hfill
\subfloat[]{
    \centering
    \includegraphics[width=0.9\columnwidth]{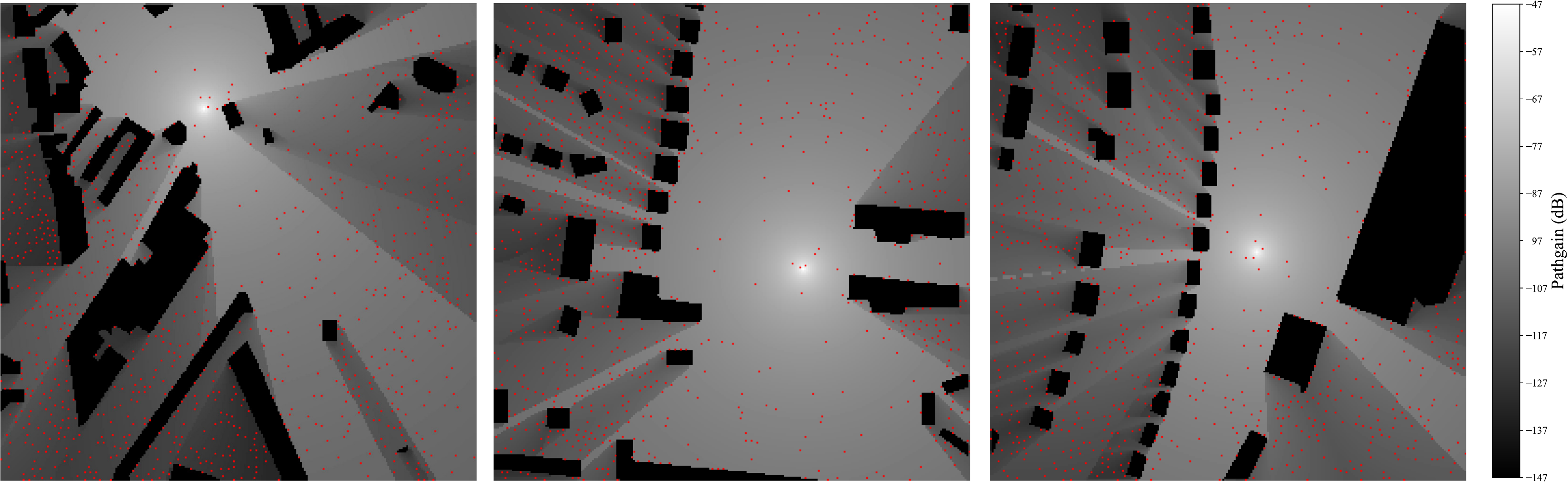}
    \label{fig_4b}
}
\caption{Spatial distribution of sampling points selected by uncertainty-aware sampling algorithm. (a) CGM prediction variance map obtained by Algorithm 1. (b) CGM prediction variance map obtained by Algorithm 2.}
\label{fig_4}
\end{figure}

For a selected candidate point $p_i$, we employ an acceptance-rejection strategy based on a predefined minimum distance $r_{\min}$. Specifically, we check whether the distance between $p_i$ and all points in the already selected set $\rm S$ is greater than $r_{\min}$. If this condition is satisfied, $p_i$ is accepted and added to the set $\rm S$. Otherwise, $p_i$ is rejected. This iterative process continues evaluating points from the candidate pool until the required number of sampling points is reached.
The above procedure is summarized in Algorithm \ref{alg3}. The spatial distributions of the observation locations obtained by the proposed method are shown in Fig. \ref{fig_4a} and Fig. \ref{fig_4b}. As observed from Fig. \ref{fig_4}, the sampling points cover the entire target area. Meanwhile, since the prediction variance is relatively large in non-line-of-sight (NLOS) regions, more sampling points are selected in these regions. In contrast, the prediction variance is smaller in line-of-sight (LOS) regions, and thus fewer sampling points are selected there. The prediction variance inside buildings is the smallest; consequently, almost no sampling points are selected inside buildings.

\begin{algorithm}
\caption{Uncertainty-Aware Sampling Algorithm}\label{alg3}
\begin{algorithmic}[1]
\Require CGM prediction variance map $\rm V$, Initial sampling points set $\rm S_0$, Number of uncertainty-aware sampling points $K$, Minimum distance between sampling points $r_{\min}$
\State Normalize the uncertainty map $\rm V$ to $\rm \tilde{V} \in [0,1]$
\State Construct the candidate pool $\rm C$ by selecting the top $q\%$ high-variance locations
\State Compute the selection probability $\rm P$ for each candidate point $p_i$ in the candidate pool $\rm C$: ${\rm P}(p_i)=\frac{{\rm \tilde{V}}_i}{\sum_{j=1}^{n}{\rm \tilde{V}}_j}$
\State Initialize $\rm S \leftarrow \rm S_0, \qquad \rm S_{\mathrm{new}} \leftarrow \emptyset$
\While{$|\rm S_{\mathrm{new}}| < K$}
    \State Draw a candidate point $p_i$ from $\rm C$ according to the
    \Statex \hspace{\algorithmicindent}
    probability $\rm P$
    \If{$\min\limits_{q \in \rm S} d(p_i,q) \ge r_{\min}$}
        \State ${\rm S}_{\mathrm{new}} \leftarrow {\rm S}_{\mathrm{new}} \cup \{p_i\}$
        \State $\rm S \leftarrow \rm S \cup \rm S_{\mathrm{new}}$
    \Else
        \State Reject $p_i$
    \EndIf
\EndWhile
\State \Return Set of all sampling points $\rm S$
\end{algorithmic}
\end{algorithm}

\subsection{Active-Learning-Based Diffusion Framework for CGM Construction}
\begin{figure*}[!t]
\centering
\includegraphics[width=\textwidth]{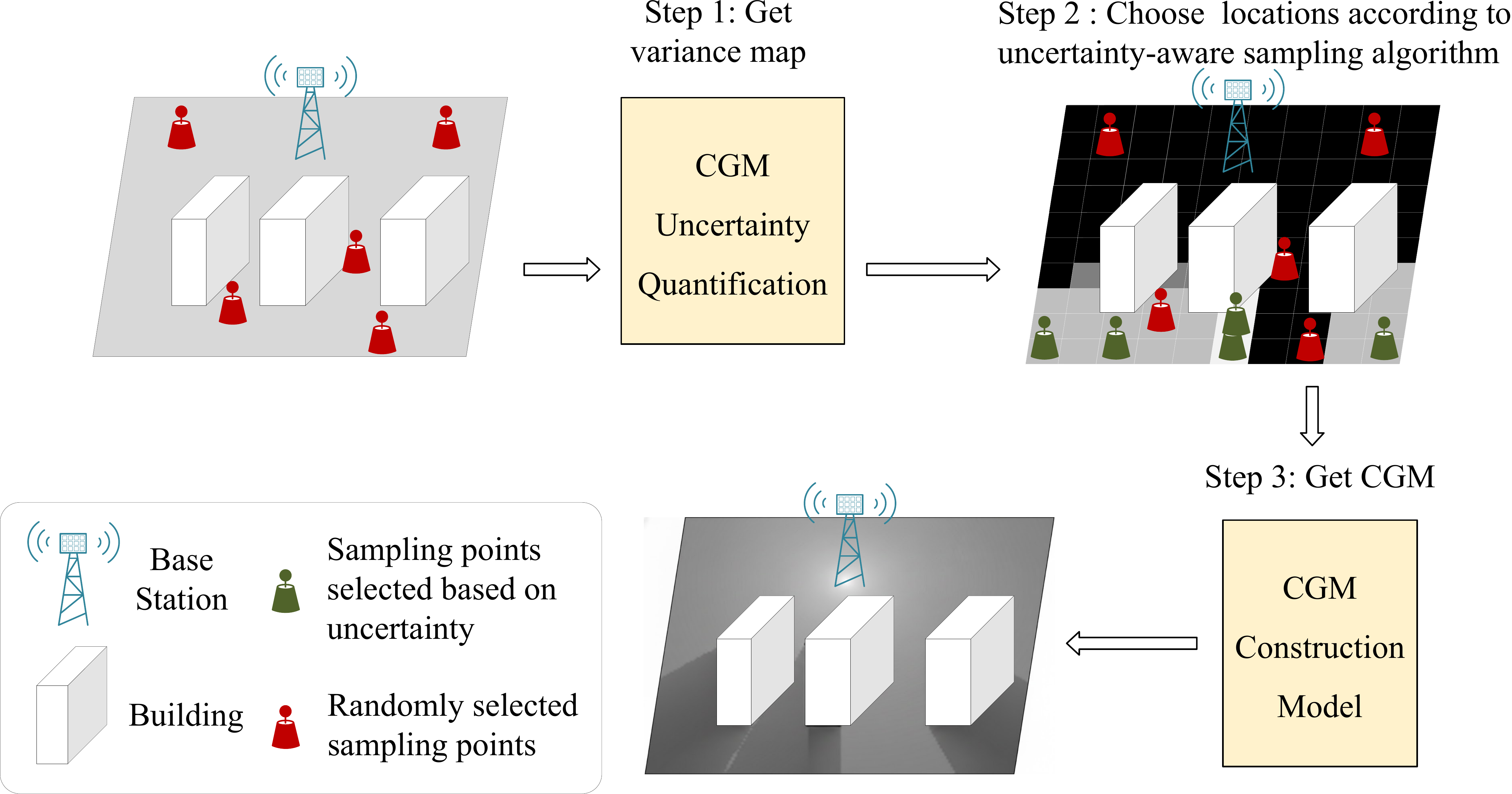}
\caption{Active-learning-based diffusion framework for CGM construction.}
\label{fig_5}
\end{figure*}

The complete workflow of the active-learning-based diffusion framework for CGM construction is illustrated in Fig. \ref{fig_5}, which consists of three steps. Given a well-trained diffusion model for CGM construction with $N$ sparse sampling points, the first step is to compute the CGM prediction variance map using either Algorithm \ref{alg1} or Algorithm \ref{alg2}. Since this step performs Bayesian inference based on the trained diffusion model, a subset of sampling points and their corresponding mask are also required as inputs. Specifically, we first randomly sample $N/2$ points to compute the CGM prediction variance map. 

The reason for setting the number of randomly sampled points to $N/2$ is as follows. Our experimental results show that, if the number of points is too small, the estimated epistemic uncertainty becomes inaccurate, and the resulting CGM prediction variance map cannot effectively guide sampling. Conversely, if an excessive number of points is allocated to this step, the remaining points for the subsequent uncertainty-aware sampling phase become insufficient, resulting in limited improvement in the final CGM construction accuracy. Balancing these two considerations, we set the number of randomly sampled points in the first step to $N/2$.

In the second step, the locations of the remaining $N/2$ points are selected according to Algorithm \ref{alg3}. 
In the third step, the measurements obtained from the random sampling in the first step and the newly acquired measurements from the uncertainty-aware sampling in the second step are combined and fed into the CGM construction model to reconstruct the complete CGM from sparse path-loss measurements.

\section{Experiment Results}\label{ER}
\subsection{Dataset}
In this paper, the proposed method is trained and evaluated on the public radio map dataset RadioMapSeer \cite{0gtx6v3022}. RadioMapSeer is constructed for urban wireless propagation scenarios and contains 700 distinct city maps collected from European urban areas. For each city map, a single transmitting base station is deployed, and ray-tracing simulations are performed using WinProp to obtain the large-scale channel fading distribution over the corresponding area. Each channel gain map is represented as a two-dimensional grid with a resolution of $256 \times 256$.
The dataset contains both static scenarios without vehicle blockage and dynamic scenarios considering the impact of vehicles, thereby reflecting the propagation characteristics under different environmental conditions. In this paper, the static and dynamic CGM data generated by the Dominant Path Model (DPM) method are used for experimental analysis. For the DPM dataset, 80 different base-station locations are randomly selected for each city map and simulated, resulting in a total of 56000 CGMs. In the experiments, the dataset is divided into training, validation, and test sets, which contain 40000, 8000, and 8000 CGMs, respectively.

\subsection{Implementation Details}
\subsubsection{Neural Network Architecture in the Diffusion Model}\label{NNADM}
The latent diffusion model adopted in this paper employs a VAE to compress the CGM with size $1 \times 256 \times 256$ into a latent feature tensor $\mathbf{z}$ with size $3 \times 64 \times 64$, and further uses the VAE decoder to decompress the generated latent feature tensor into the final generated result. 

The backbone network of the diffusion model is implemented using a U-Net architecture. The encoder part of the U-Net consists of four downsampling blocks, and the corresponding decoder part consists of four upsampling blocks. Both the downsampling and upsampling blocks contain residual convolutional units for downsampling or upsampling, as well as cross-attention units. 

The condition encoder is implemented using a Swin-T network. 
When the input condition is encoded by Swin-T, the output feature tensor is used in two ways. On the one hand, it is concatenated with the latent feature tensor $\mathbf{z}$ along the channel dimension and fed into the U-Net as input. On the other hand, it is transformed by convolutional layers and then injected into the U-Net backbone through the aforementioned cross-attention mechanism.

\subsubsection{Hyperparameter Settings}
During training, we adopt the standard DDPM training paradigm. The number of diffusion time steps is set to 1000, and the prediction target is set to the $\rm pred-x_0$ mode. The hyperparameters $\alpha_t$ and $\beta_t$ are determined according to the cosine noise schedule. The training process consists of two stages. In the first stage, the VAE is trained using the AdamW optimizer. To reduce GPU memory consumption during training, gradient accumulation is adopted. Specifically, the mini-batch size is set to 8, and gradients are accumulated over two iterations before performing backpropagation and parameter update, resulting in an effective batch size of 16. The total number of training steps is set to 150,000, corresponding to 60 effective training epochs. The initial learning rate $l_0$ is set to $1\times 10^{-5}$, and the minimum learning rate $l_{min}$ is set to $1\times 10^{-6}$. 
In the second stage, the backbone U-Net is trained. The optimizer and learning-rate settings are the same as those used for VAE training. Gradient accumulation is also adopted in this stage. Specifically, the mini-batch size is set to 6, and gradients are accumulated over 12 iterations before backpropagation and parameter update, resulting in an effective batch size of 72. The total number of training steps is set to 100,000, corresponding to 180 effective training epochs. 

In the implementation of Algorithms \ref{alg1} and \ref{alg2},
the parameter posterior is computed using the laplace-torch library \cite{daxberger2021laplace}, which provides an efficient implementation for the complex covariance matrix computation involved in the Laplace approximation for deep learning models.

Both algorithms assume that the parameter prior $p(\theta)$ follows a standard Gaussian distribution with zero mean and an identity covariance matrix.
In both Algorithms \ref{alg1} and \ref{alg2}, the number of DDIM sampling steps is set to 5, and the number of Monte Carlo samples is set to 16. 
The main difference between the two algorithms is that Algorithm \ref{alg1} computes the full parameter posterior covariance matrix, whereas Algorithm \ref{alg2} computes only the diagonal elements of the parameter posterior covariance matrix. 
In addition, as described in Section~\ref{MRQU}, Algorithm \ref{alg2} adopts a partially parallelized computation strategy, where four sets of last-layer network parameters are loaded simultaneously to improve computational efficiency. For final CGM generation, the DDIM sampling algorithm is also adopted, with the number of sampling steps set to 5. In Algorithm \ref{alg3}, we select the locations with the top 80\% highest variances as the sampling candidate pool $C$, and constrain the minimum distance $r_{\min}$ between any two sampling points to be 3.

In addition, considering that the measurements in practical scenarios may be corrupted by noise, additive white Gaussian noise (AWGN) is added to the input sparse observations during both the training and generation stages. The standard deviation of AWGN is set to $\sigma = 0.03$.

\subsection{Performance Comparison}
\subsubsection{Baseline Method}
In this paper, three random-sampling-based CGM construction methods are adopted as baseline methods, including RadioUNet~\cite{9354041}, RME-GAN~\cite{10130091}, and the latent diffusion model described in Section~\ref{NNADM}. For a fair comparison, all baseline methods are trained with the same number of epochs and use sparse observations obtained by random sampling as inputs.


Furthermore, the primary objective of this paper is to introduce an active learning mechanism into an existing diffusion-based CGM construction model. To validate the effectiveness of the proposed active-learning-based diffusion framework, we use the same latent diffusion model as the baseline method for CGM reconstruction. By comparing the reconstruction performance of different approaches with the same number of sampling points, we evaluate the improvement achieved by the proposed method in terms of CGM construction accuracy.
\subsubsection{Evaluation Metric}
In the CGM construction problem, the reconstruction accuracy is of primary interest. Therefore, we adopt three evaluation metrics derived from the mean squared error (MSE), namely the normalized mean squared error (NMSE), the root mean squared error (RMSE), and the peak signal-to-noise ratio (PSNR).

NMSE is obtained by normalizing the MSE. Let $\hat{X}$ denote the CGM reconstructed by the model, $X$ denote the ground-truth CGM, and $N$ denote the number of samples in the test set. The NMSE is then computed as follows:
\begin{equation}
    \label{eq_45}
    {\rm NMSE} = \frac{1}{N} \sum_{n=1}^{N} \frac{\sum_{i=1}^{H} \sum_{j=1}^{W} \left| \hat{X}_{n}(i,j) - X_n(i,j) \right|^2}{\sum_{i=1}^{H} \sum_{j=1}^{W} \left| X_n(i,j) \right|^2}.
\end{equation}

RMSE is obtained by taking the square root of MSE and therefore has the same unit as the original data, making it convenient for intuitively evaluating the error magnitude. It is computed as follows:
\begin{equation}
    \label{eq_46}
    {\rm RMSE} = \sqrt{ \frac{1}{N} \sum_{n=1}^{N} \frac{1}{H \times W} \sum_{i=1}^{H} \sum_{j=1}^{W} \left| \hat{X}_n(i,j) - X_n(i,j) \right|^2 }.
\end{equation}

PSNR represents the reconstruction error in a logarithmic form from the perspective of the signal-to-noise ratio. It is computed as follows, where $\rm MAX$ denotes the peak value of the CGM. Therefore, unlike NMSE and RMSE, a larger PSNR indicates higher reconstruction accuracy.
\begin{equation}
    \label{eq_47}
    {\rm PSNR} =\frac{1}{N} \sum_{n=1}^N 10 \lg \left(\frac{\rm MAX^2}{\rm MSE_n}\right).
\end{equation}

\subsubsection{CGM Prediction Variance Map}
\begin{figure}[!htbp]
    \centering
    \subfloat[\label{fig6_a}]{
        \includegraphics[width=0.305\columnwidth]{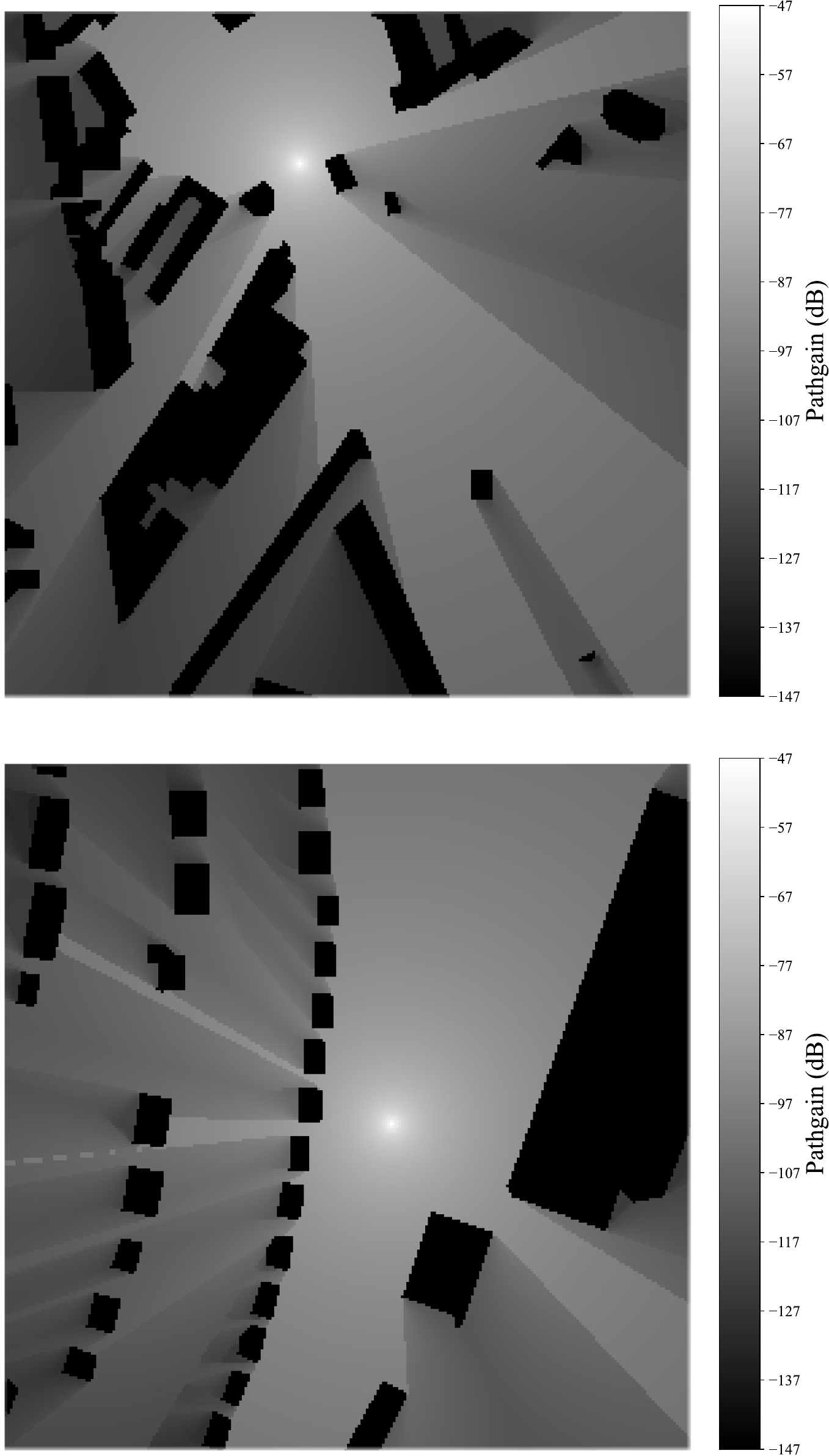}
    }
    \hfill
    \subfloat[\label{fig6_b}]{
        \includegraphics[width=0.305\columnwidth]{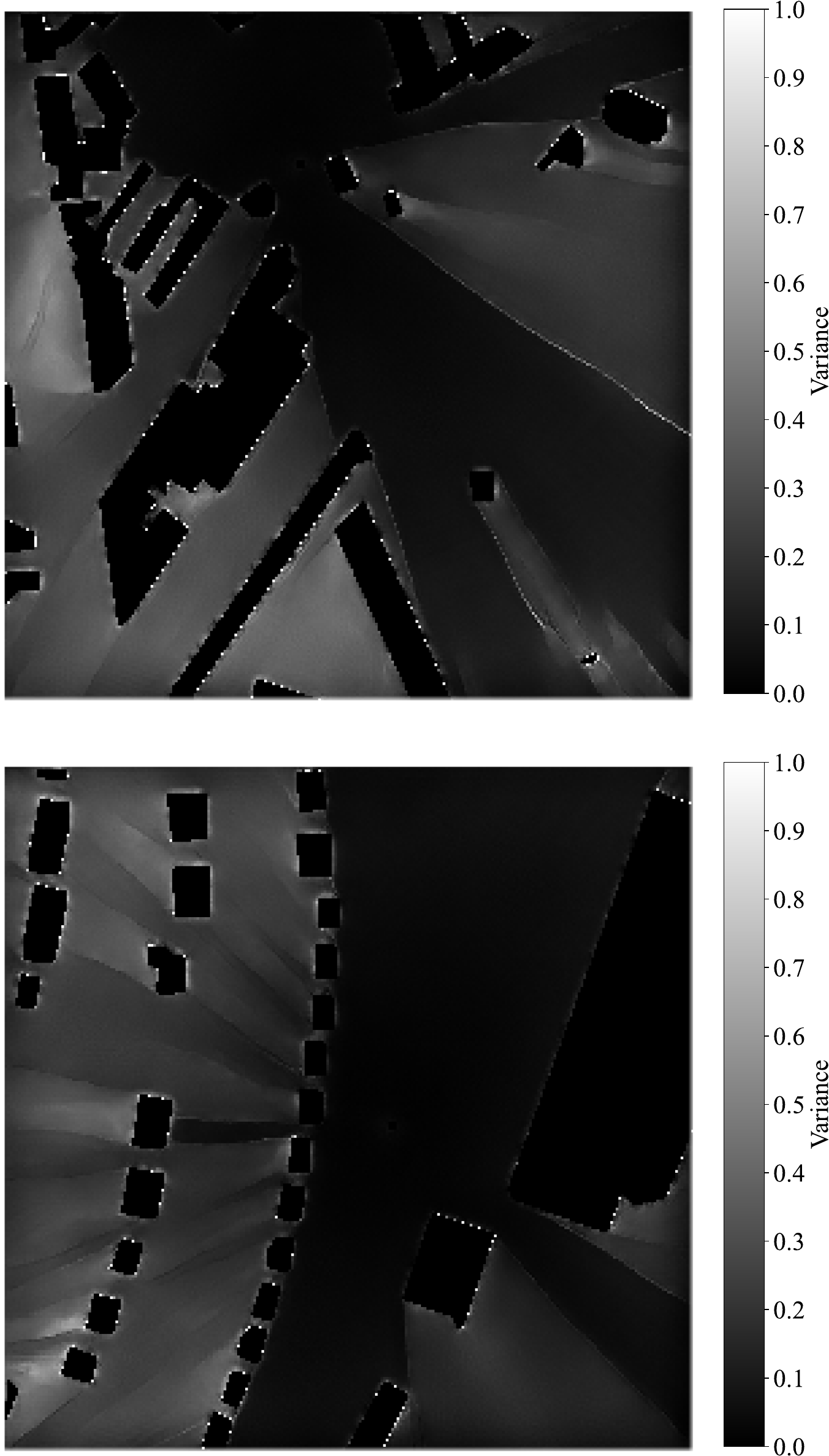}
    }
    \hfill
    \subfloat[\label{fig6_c}]{
        \includegraphics[width=0.305\columnwidth]{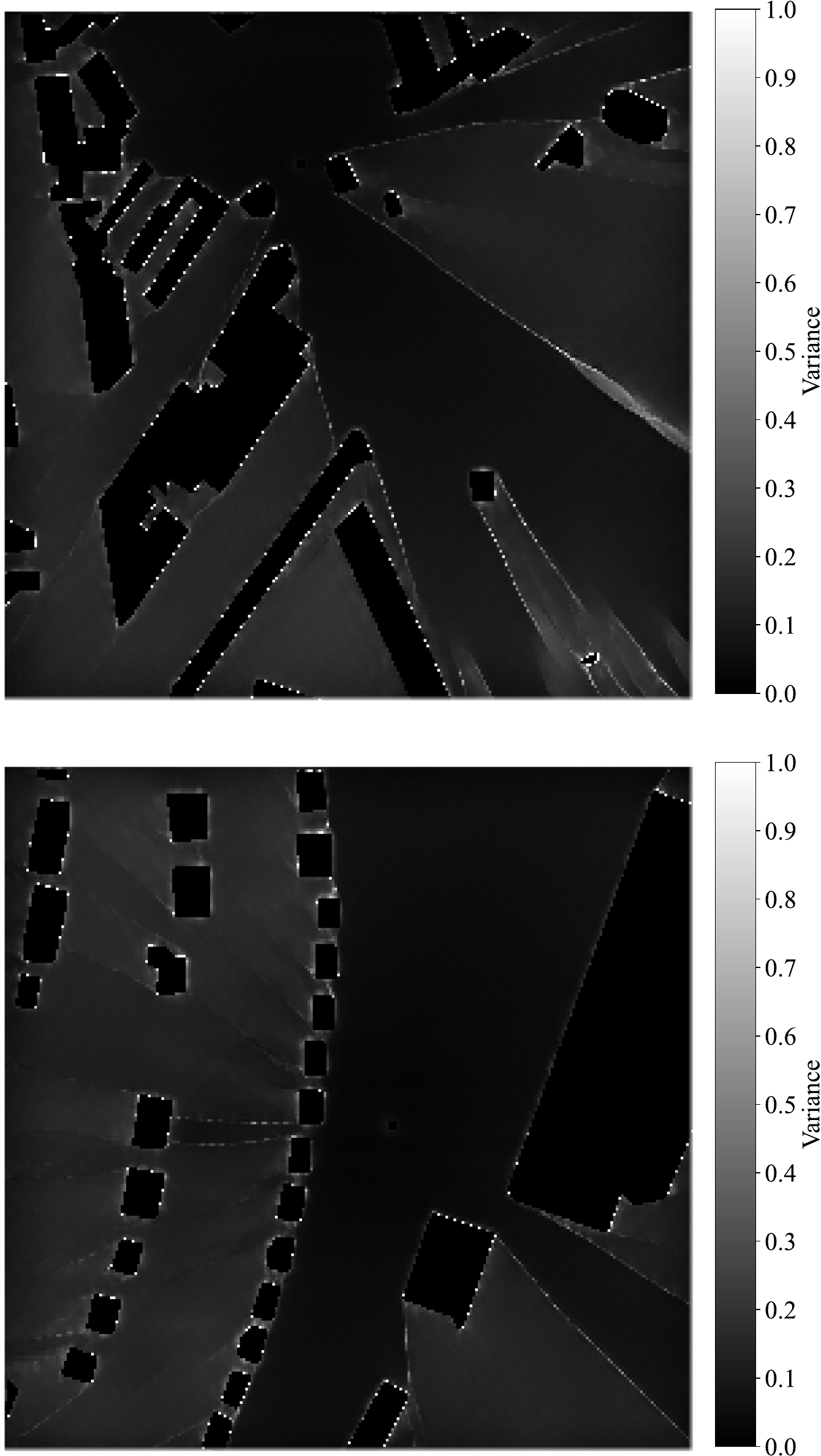}
    }
    \caption{Static CGM. (a) Ground truth CGMs. (b) Variance maps using Algorithm 1. (c) Variance maps using Algorithm 2}
    \label{fig6}
\end{figure}

\begin{figure}[!htbp]
    \centering
    \subfloat[\label{fig7_a}]{
        \includegraphics[width=0.305\columnwidth]{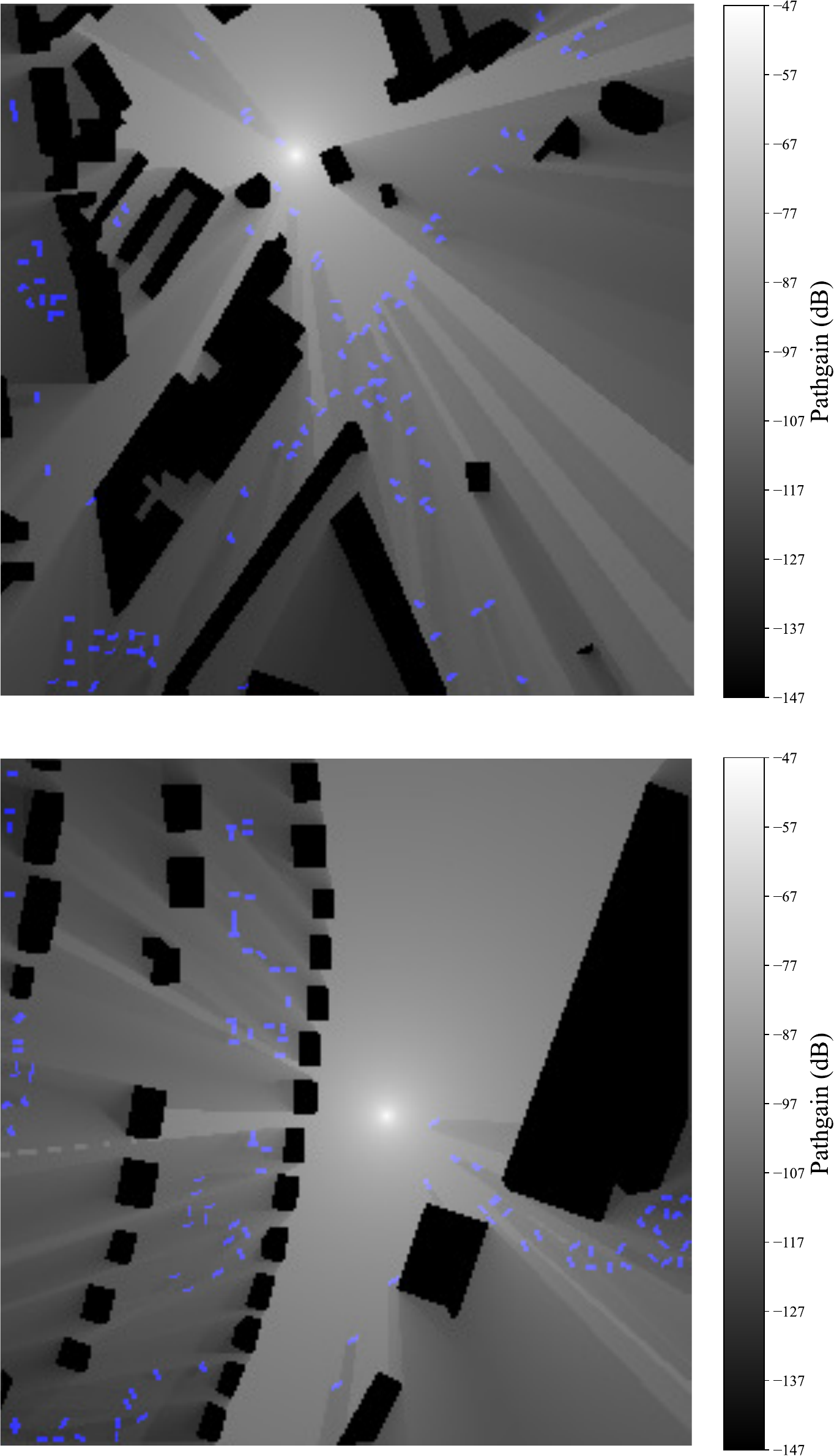}
    }
    \hfill
    \subfloat[\label{fig7_b}]{
        \includegraphics[width=0.305\columnwidth]{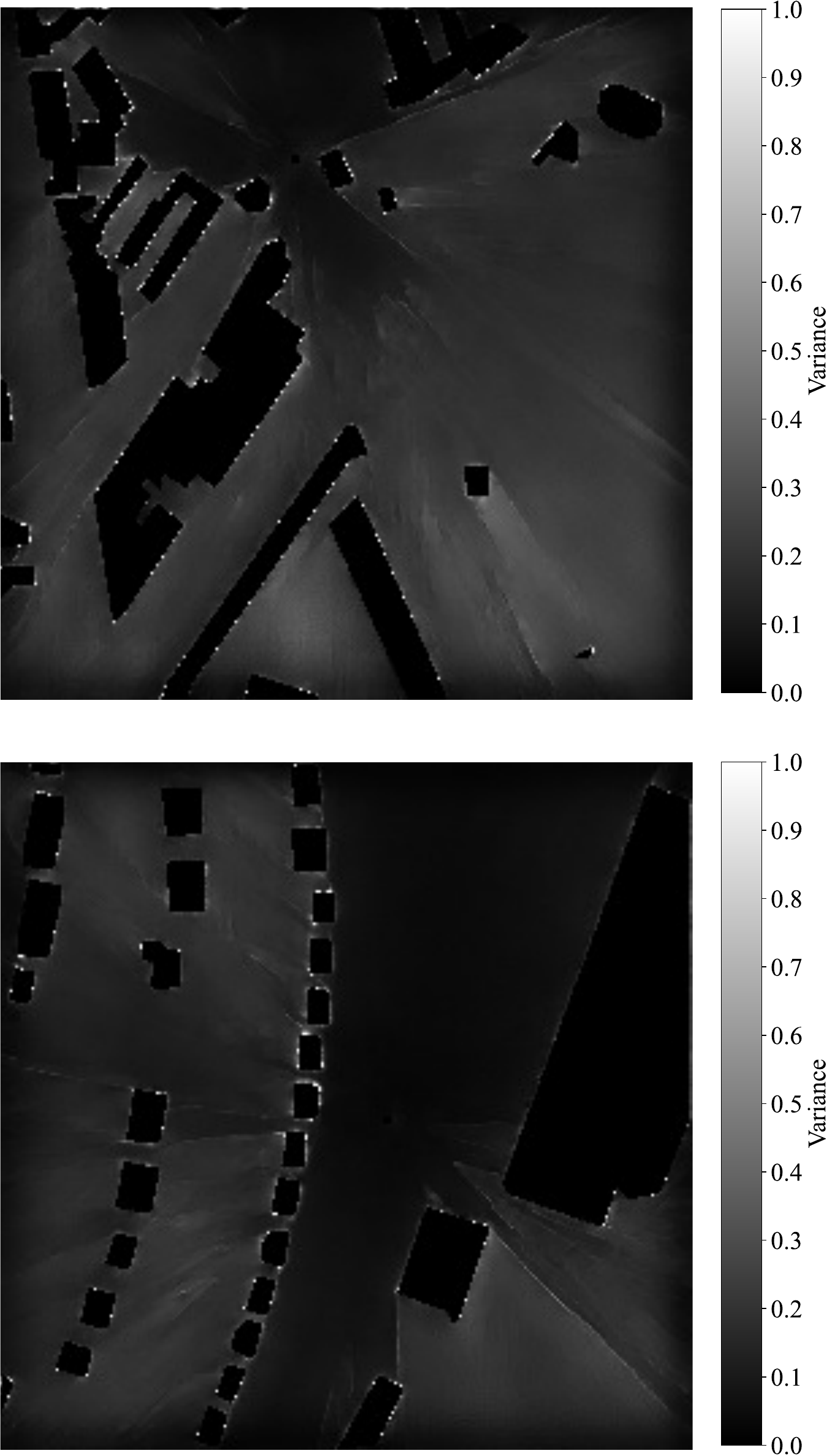}
    }
    \hfill
    \subfloat[\label{fig7_c}]{
        \includegraphics[width=0.305\columnwidth]{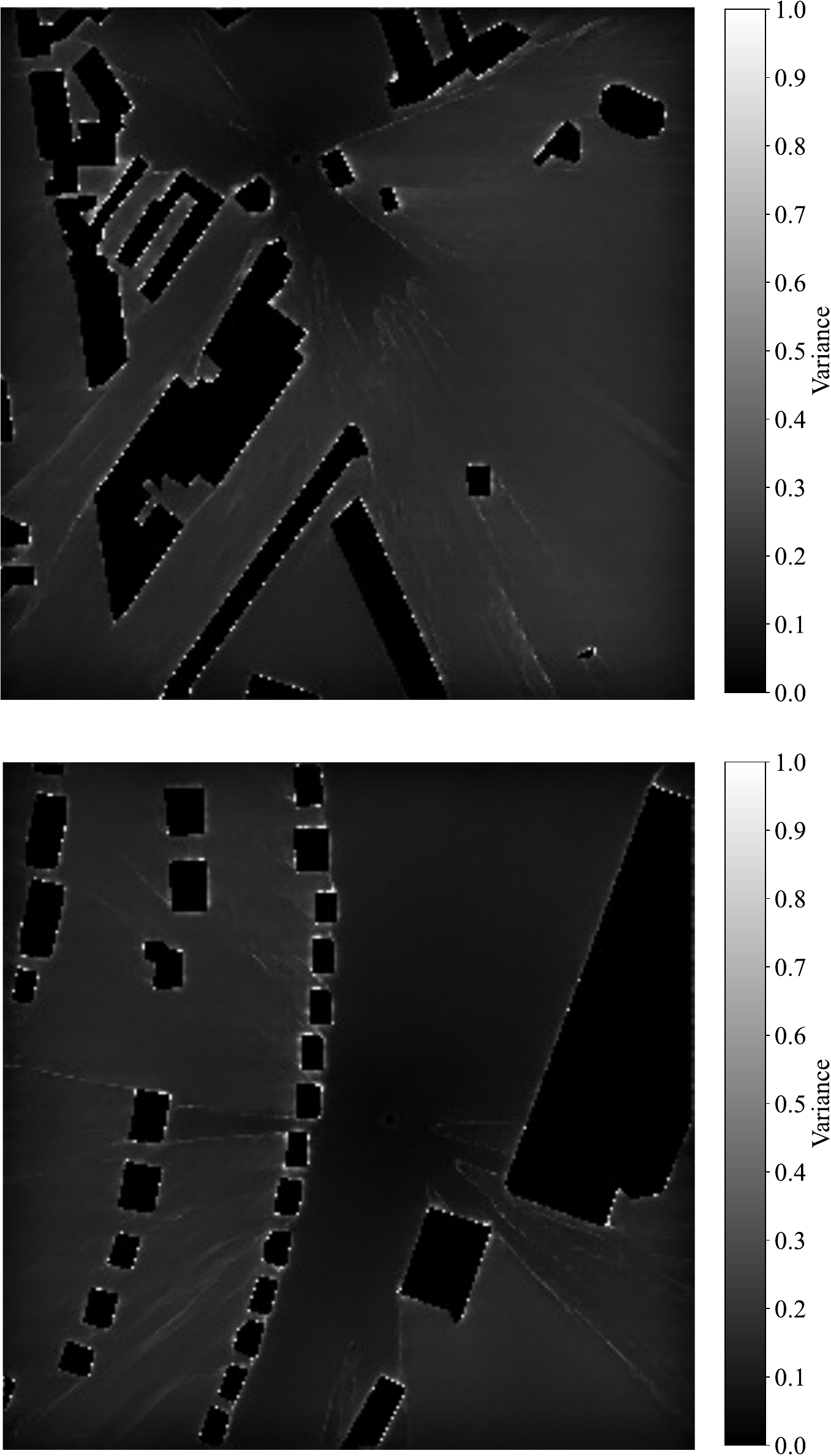}
    }
    \caption{Dynamic CGM. (a) Ground truth CGMs. (b) Variance maps using Algorithm 1. (c) Variance maps using Algorithm 2}
    \label{fig7}
\end{figure}

Figs. \ref{fig6} and \ref{fig7} illustrate the CGM prediction variance maps derived from Algorithm \ref{alg1} and Algorithm \ref{alg2} under static and dynamic scenarios, respectively. We uniformly represent both the CGMs and the variance maps as grayscale images, where darker pixels in the CGMs denote higher path loss, and brighter pixels in the variance maps indicate greater prediction uncertainty. As shown in Fig. \ref{fig6}, the variance maps from both algorithms share consistent spatial distribution patterns. 

According to Section~\ref{LA}, this variance quantifies the model uncertainty stemming from the limitations of its predictive capability. In the CGM construction task, the received signal strength inside building interiors is identically zero, and the model's prediction uncertainty within these areas is minimal. In LOS regions, the spatial variation of large-scale fading adheres to the log-distance path loss model, which likewise results in relatively low prediction uncertainty. Conversely, large-scale fading undergoes abrupt transitions at building edges and the boundaries between LOS and NLOS regions. These areas are the most challenging to predict in practice, consequently yielding the highest uncertainty. Furthermore, in deep NLOS regions, severe blockages restrict the availability of informative observations for the model, thereby further elevating the prediction uncertainty. The spatial distribution characteristics of the CGM prediction variance map shown in Fig. \ref{fig6_b} and \ref{fig6_c} are in highly consistent agreement with the aforementioned physical analysis, substantiating the physical correctness of our obtained variance maps and demonstrating their utility in guiding uncertainty-aware sampling strategies.

In the dynamic CGMs shown in Fig. \ref{fig7}, the blue points represent vehicles. Compared to the static CGMs, the primary distinction is the introduction of these vehicles as additional moving obstacles. Consequently, the spatial distribution patterns in the dynamic CGMs become significantly more complex and challenging to predict. Nevertheless, the spatial variation of the model's predictive capability in the dynamic scenarios remains consistent with that in the static cases. Specifically, locations experiencing abrupt transitions in path loss and areas subjected to severe obstruction pose greater prediction difficulties, which should correspondingly exhibit larger prediction variances. The CGM prediction variance maps derived from both algorithms in Fig. \ref{fig7_b} and \ref{fig7_c} successfully reflect these characteristics.

\begin{figure}[!htbp]
    \centering
    \subfloat[\label{fig8_a}]{
        \includegraphics[width=0.45\columnwidth]{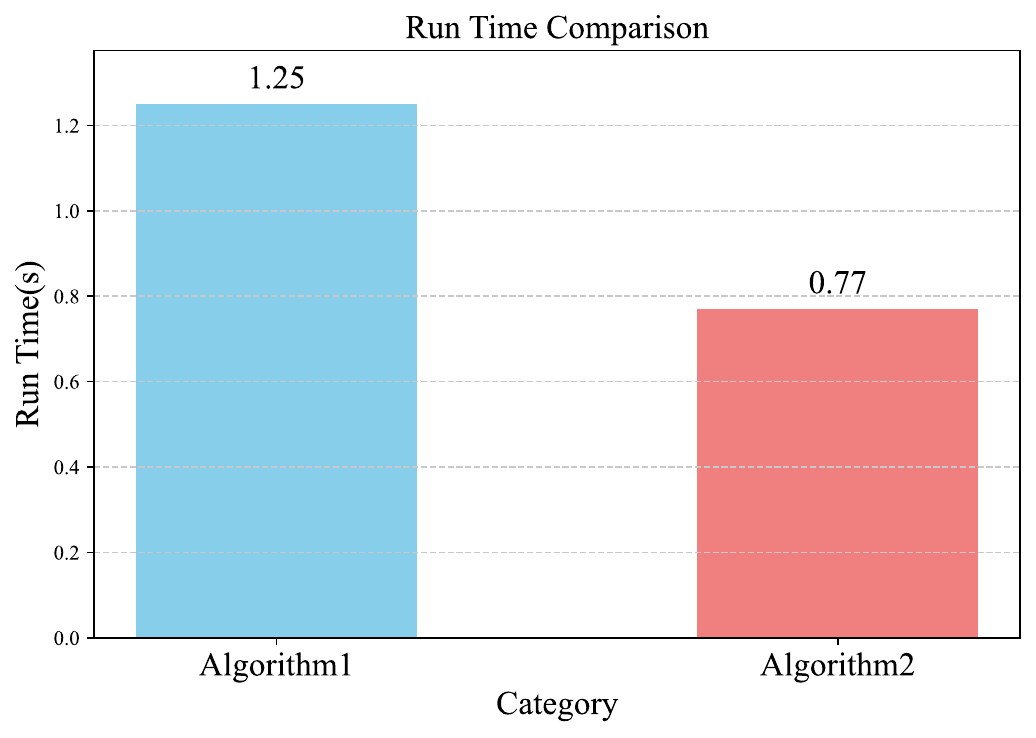}
    }
    \hfill
    \subfloat[\label{fig8_b}]{
        \includegraphics[width=0.45\columnwidth]{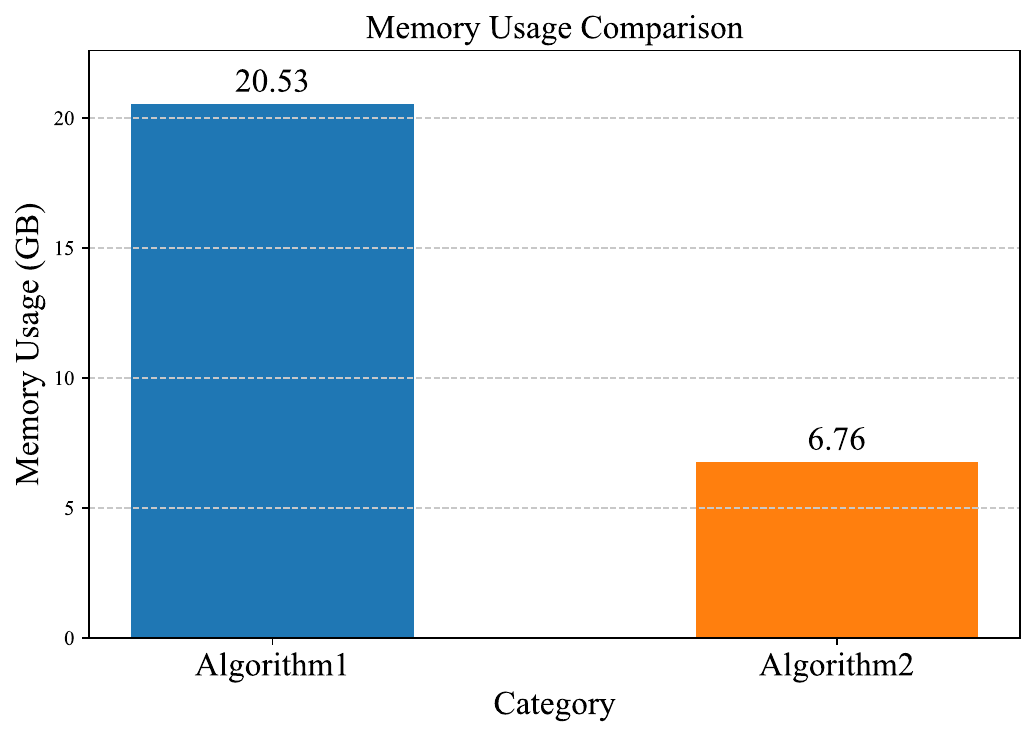}
    }
    \caption{Comparison of computational time and GPU memory consumption. (a) Computational time. (b) GPU memory consumption.}
    \label{fig8}
\end{figure}

Furthermore, we compare the computational time and GPU memory consumption required by the two algorithms for generating the variance maps, 
as illustrated in Figs. \ref{fig8_a} and \ref{fig8_b}.
As can be observed, Algorithm \ref{alg1} requires 1.25 s and consumes 13.53 GB of GPU memory. In contrast, Algorithm \ref{alg2} takes only 0.77 s and consumes 6.76 GB of GPU memory. These results demonstrate that Algorithm \ref{alg2} is a more efficient approach for computing the CGM prediction variance maps.

\subsubsection{Comparison with Baseline Methods}
\begin{figure*}[!htbp]
    \centering
    \subfloat[\label{fig9_a}]{
        \includegraphics[width=0.31\textwidth]{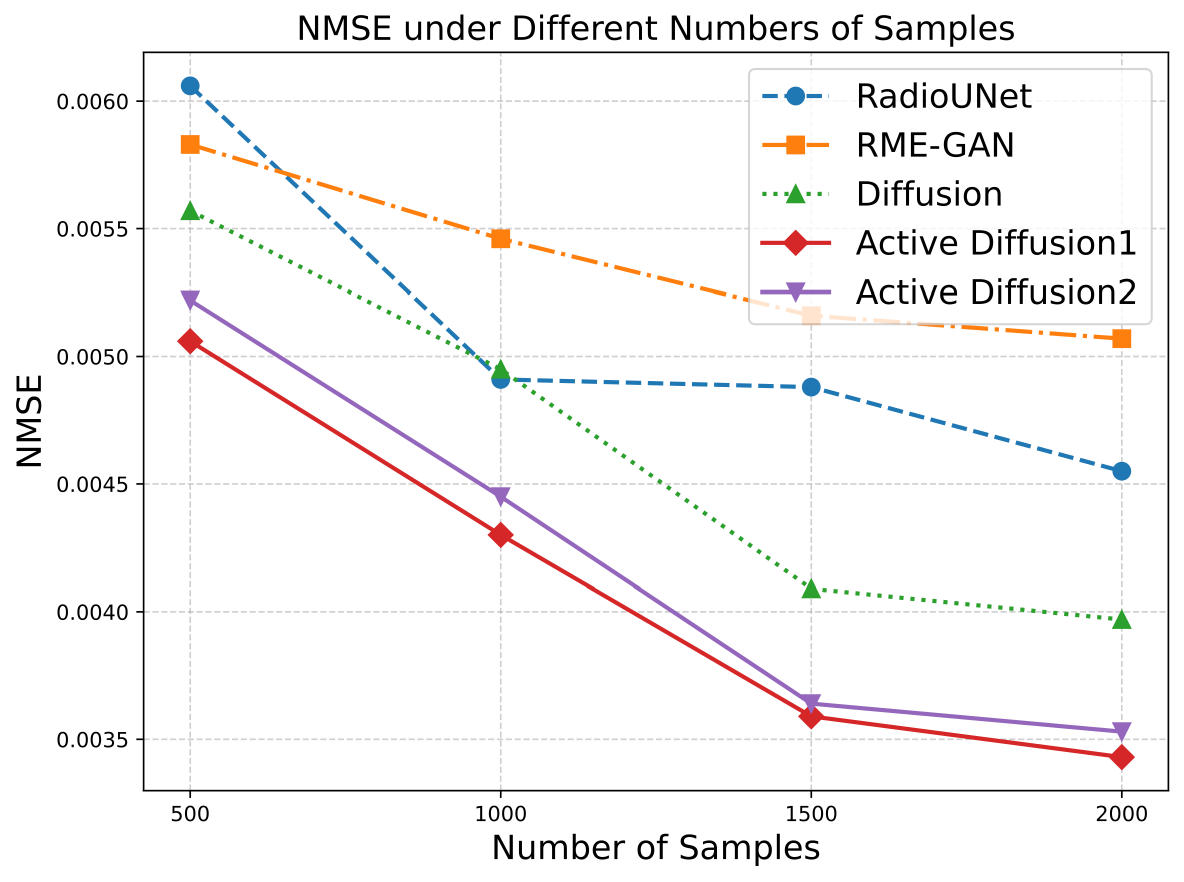}
    }
    \hfill
    \subfloat[\label{fig9_b}]{
        \includegraphics[width=0.31\textwidth]{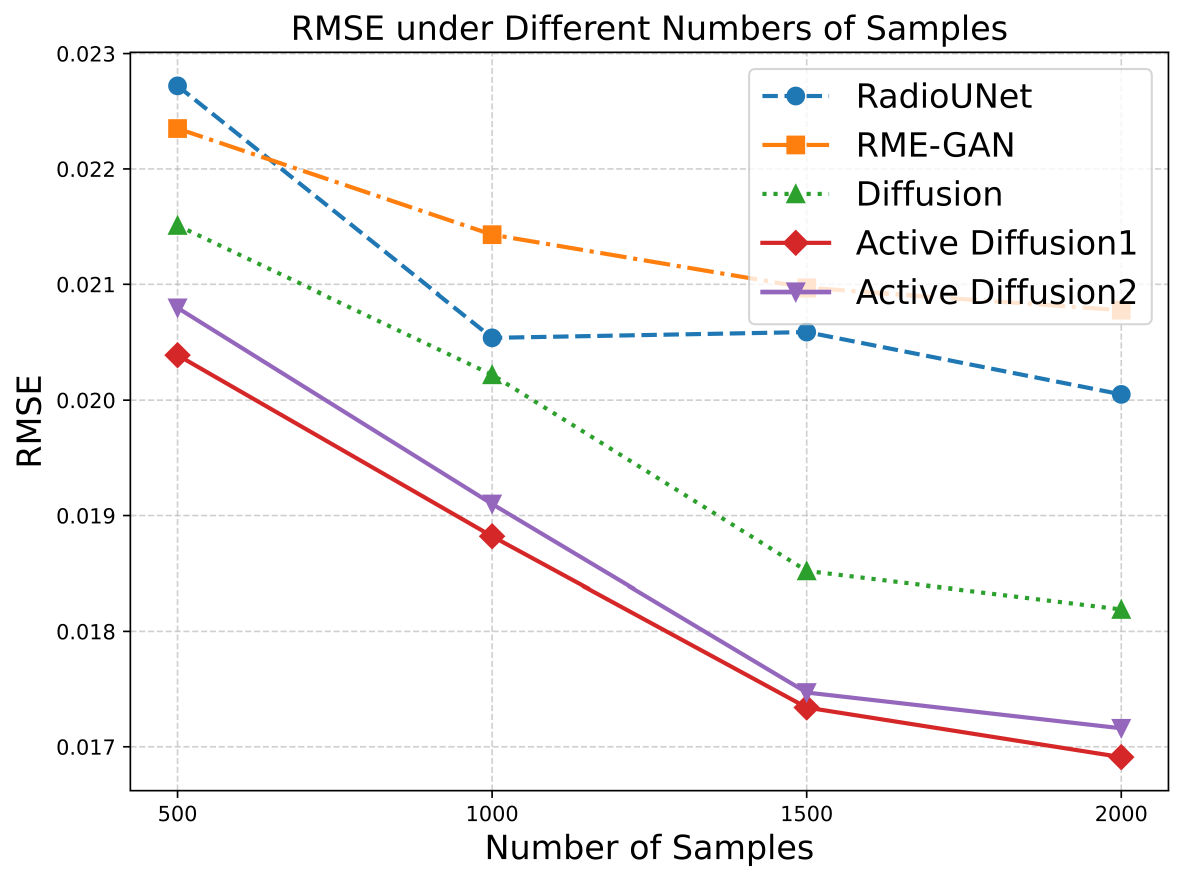}
    }
    \hfill
    \subfloat[\label{fig9_c}]{
        \includegraphics[width=0.31\textwidth]{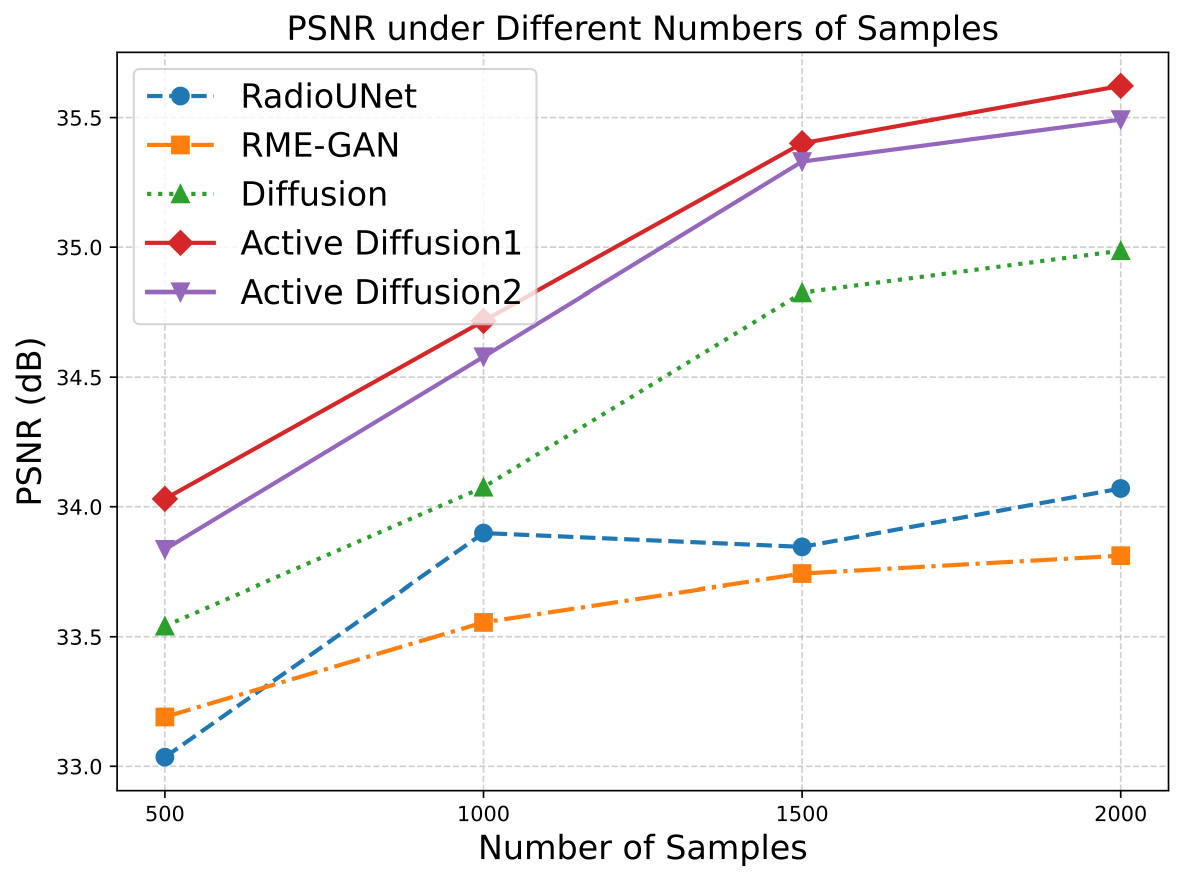}
    }
    \caption{Static CGM. (A) NMSE. (B) RMSE. (C) PSNR.}
    \label{fig9}
\end{figure*}

Tables~\ref{tab1} and~\ref{tab2} report the construction performance of five methods on the static and dynamic CGM datasets, respectively. 
Diffusion denotes the use of the latent diffusion model described in Section~\ref{NNADM} for CGM construction.
Active Diffusion 1 and Active Diffusion 2 denote the proposed methods using the element-wise epistemic uncertainty computed by Algorithm \ref{alg1} and Algorithm \ref{alg2}, respectively, while employing the same diffusion model as Diffusion for CGM construction. The experiments are conducted with 500, 1000, 1500, and 2000 sampling points. 
The variations of NMSE, RMSE, and PSNR with respect to the number of sampling points on the static CGM dataset are shown in Fig.~\ref{fig9_a}--\ref{fig9_c}, while the corresponding results on the dynamic CGM dataset are shown in Fig.~\ref{fig11_a}--\ref{fig11_c}.

For the static radio map dataset, the NMSE and RMSE of all five methods generally decrease as the number of sampling points increases, while PSNR exhibits an overall increasing trend. This indicates that additional sparse observations provide more sufficient measurement constraints for CGM construction, thereby improving reconstruction accuracy. Across varying numbers of sampling points, the diffusion model consistently outperforms RadioUNet and RME-GAN, which suggests that the diffusion model is better suited to sparse and noisy observation conditions.

\begin{table}[!t]
    \begin{threeparttable}
        \caption{Construction Accuracy Comparison on the Static CGM Dataset}
        \label{tab1}
        \centering
        \footnotesize
        \setlength{\tabcolsep}{5pt}
        \begin{tabular*}{\textwidth}{@{\extracolsep{\fill}}clccc@{}}
            \toprule
            Sampling Points & Method & NMSE & RMSE & PSNR (dB) \\
            \midrule
            \multirow{5}{*}{500}
            & RadioUNet & 0.00606 & 0.02272 & 33.03537 \\
            & RME-GAN & 0.00583 & 0.02235 & 33.19037 \\
            & Diffusion & 0.00557 & 0.02151 & 33.54112 \\
            & Active Diffusion 1 & \textbf{0.00506} & \textbf{0.02039} & \textbf{34.03009} \\
            & Active Diffusion 2 & \underline{0.00522} & \underline{0.02080} & \underline{33.83537} \\
            \addlinespace[1pt]
            \midrule
            \multirow{5}{*}{1000}
            & RadioUNet & 0.00491 & 0.02054 & 33.89828 \\
            & RME-GAN & 0.00546 & 0.02143 & 33.55541 \\
            & Diffusion & 0.00495 & 0.02022 & 34.07534 \\
            & Active Diffusion 1 & \textbf{0.00430} & \textbf{0.01882} & \textbf{34.71696} \\
            & Active Diffusion 2 & \underline{0.00445} & \underline{0.01910} & \underline{34.57868} \\
            \addlinespace[1pt]
            \midrule
            \multirow{5}{*}{1500}
            & RadioUNet & 0.00488 & 0.02059 & 33.84501 \\
            & RME-GAN & 0.00516 & 0.02097 & 33.74205 \\
            & Diffusion & 0.00409 & 0.01852 & 34.82499 \\
            & Active Diffusion 1 & \textbf{0.00359} & \textbf{0.01734} & \textbf{35.40086} \\
            & Active Diffusion 2 & \underline{0.00364} & \underline{0.01747} & \underline{35.330265} \\
            \addlinespace[1pt]
            \midrule
            \multirow{5}{*}{2000}
            & RadioUNet & 0.00455 & 0.02005 & 34.07026 \\
            & RME-GAN & 0.00507 & 0.02078 & 33.81097 \\
            & Diffusion & 0.00397 & 0.01819 & 34.98610 \\
            & Active Diffusion 1 & \textbf{0.00343} & \textbf{0.01691} & \textbf{35.62244} \\
            & Active Diffusion 2 & \underline{0.00353} & \underline{0.01716} & \underline{35.49214} \\
            \bottomrule
        \end{tabular*}
        \begin{tablenotes}
            \footnotesize
            \item[] \textbf{Note:} Bold and underlined values indicate the best and second-best performance, respectively.
        \end{tablenotes}
    \end{threeparttable}
\end{table}

\begin{figure}[!htbp]
\centering
\includegraphics[width=\columnwidth]{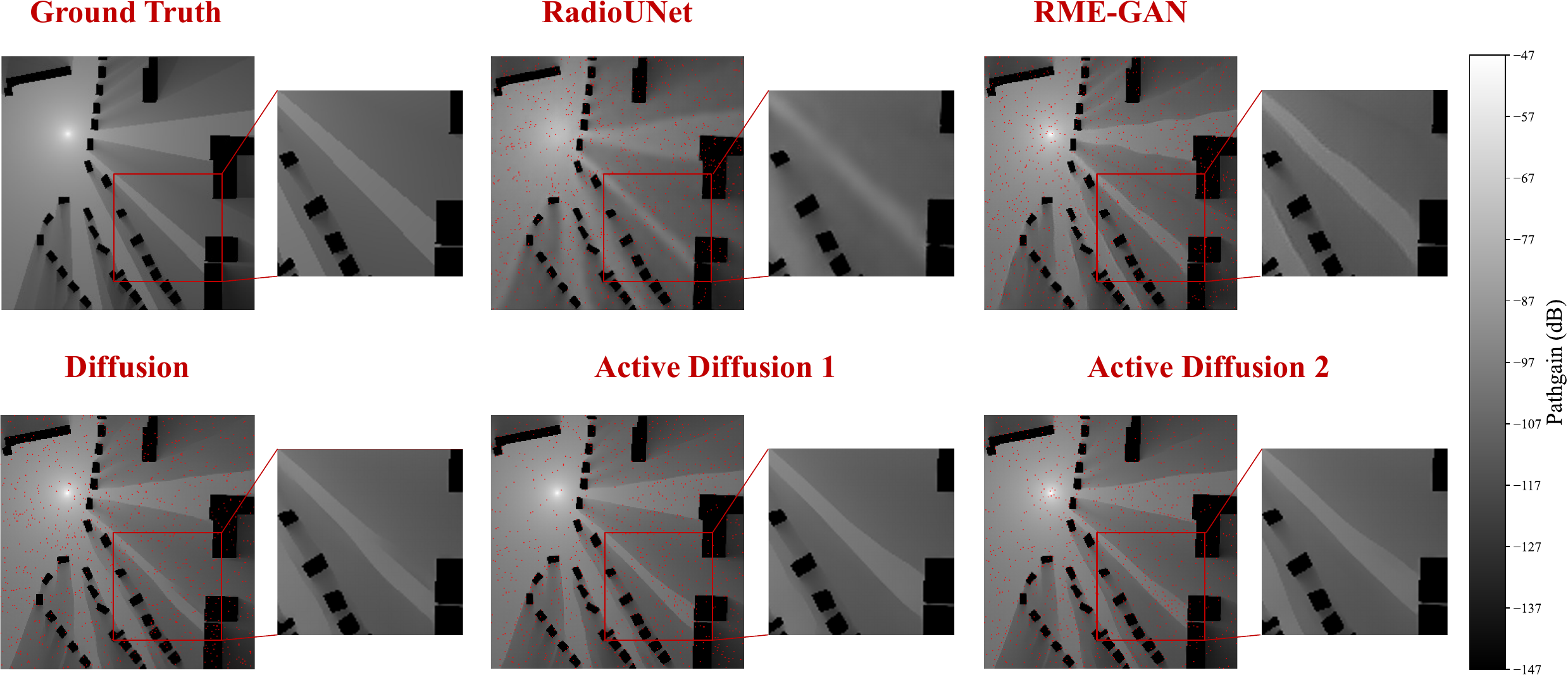}
\caption{Comparison of static CGMs constructed by different models.}
\label{fig10}
\end{figure}

With 500 sampling points, the NMSE values achieved by the proposed Active Diffusion 1 and Active Diffusion 2 are reduced to 0.00506 and 0.00522, respectively. Compared with the original Diffusion method, Active Diffusion 1 reduces NMSE by approximately 9.16\%, decreases RMSE from 0.02151 to 0.02039, and improves PSNR from 33.54~dB to 34.03~dB. As the number of sampling points increases to 1000, 1500, and 2000, the advantages of the proposed active learning mechanism are still maintained. On the static dataset, Active Diffusion 1 achieves the best performance across all four sampling budgets. Relative to the original Diffusion method, Active Diffusion 1 reduces NMSE by approximately 13.13\%, 12.22\%, and 13.60\% with 1000, 1500, and 2000 sampling points, respectively, and improves PSNR by 0.64~dB, 0.58~dB, and 0.64~dB. Similarly, Active Diffusion 2 reduces NMSE by approximately 10.10\%, 11.00\%, and 11.08\%, and improves PSNR by 0.50~dB, 0.51~dB, and 0.51~dB across these same sampling budgets. These results demonstrate that active learning mechanism can preferentially select points that are more informative for reconstruction, thereby maximizing the utilization efficiency of sparse observations. Although Algorithm \ref{alg1} incurs higher computational complexity than Algorithm \ref{alg2}, it computes the full covariance matrix of the parameter posterior distribution. Therefore, it can characterize the correlations among model parameters more accurately, leading to more reliable epistemic uncertainty estimates. The sampling points selected according to this uncertainty quantification typically carry higher information value and provide more effective observation constraints for subsequent CGM reconstruction. Consequently, Active Diffusion 1 generally provides higher CGM construction accuracy than Active Diffusion 2.

\begin{figure*}[!htbp]
    \centering
    \subfloat[\label{fig11_a}]{
        \includegraphics[width=0.31\textwidth]{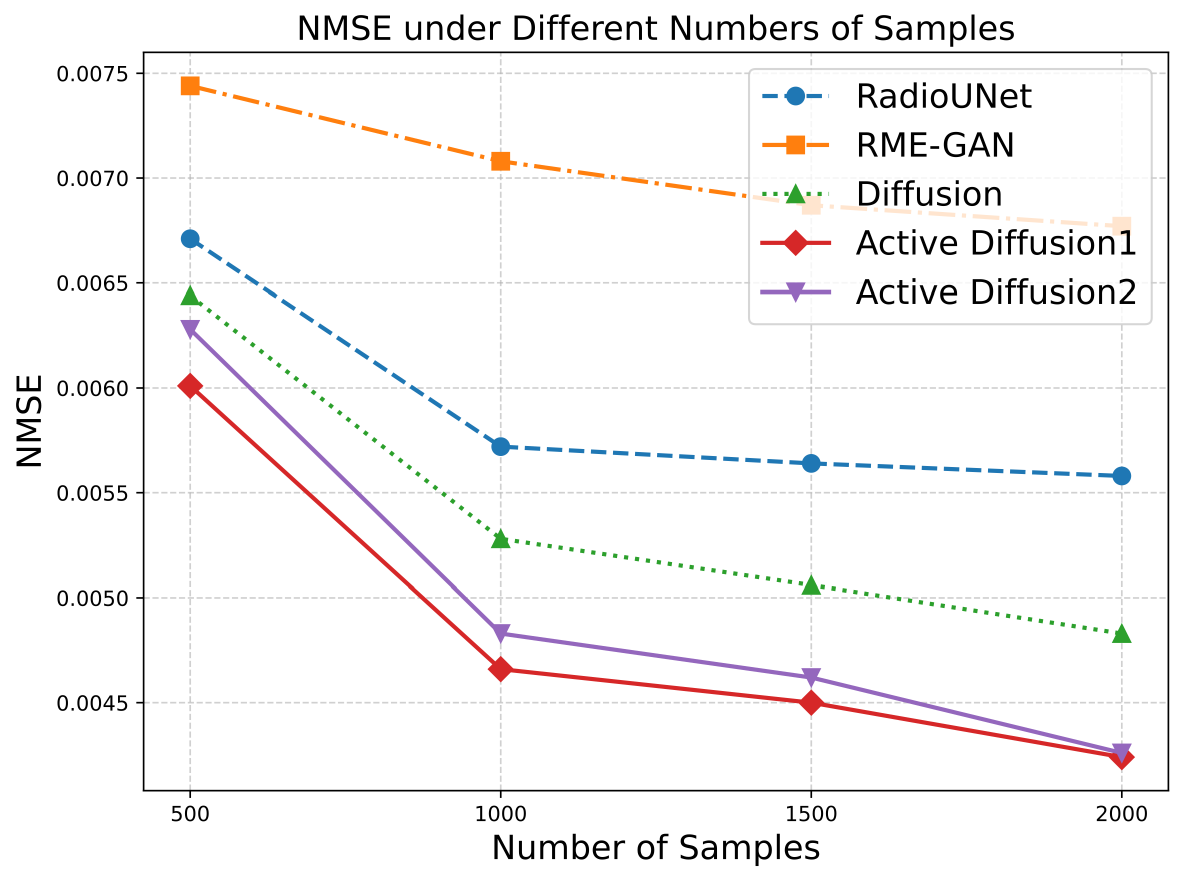}
    }
    \hfill
    \subfloat[\label{fig11_b}]{
        \includegraphics[width=0.31\textwidth]{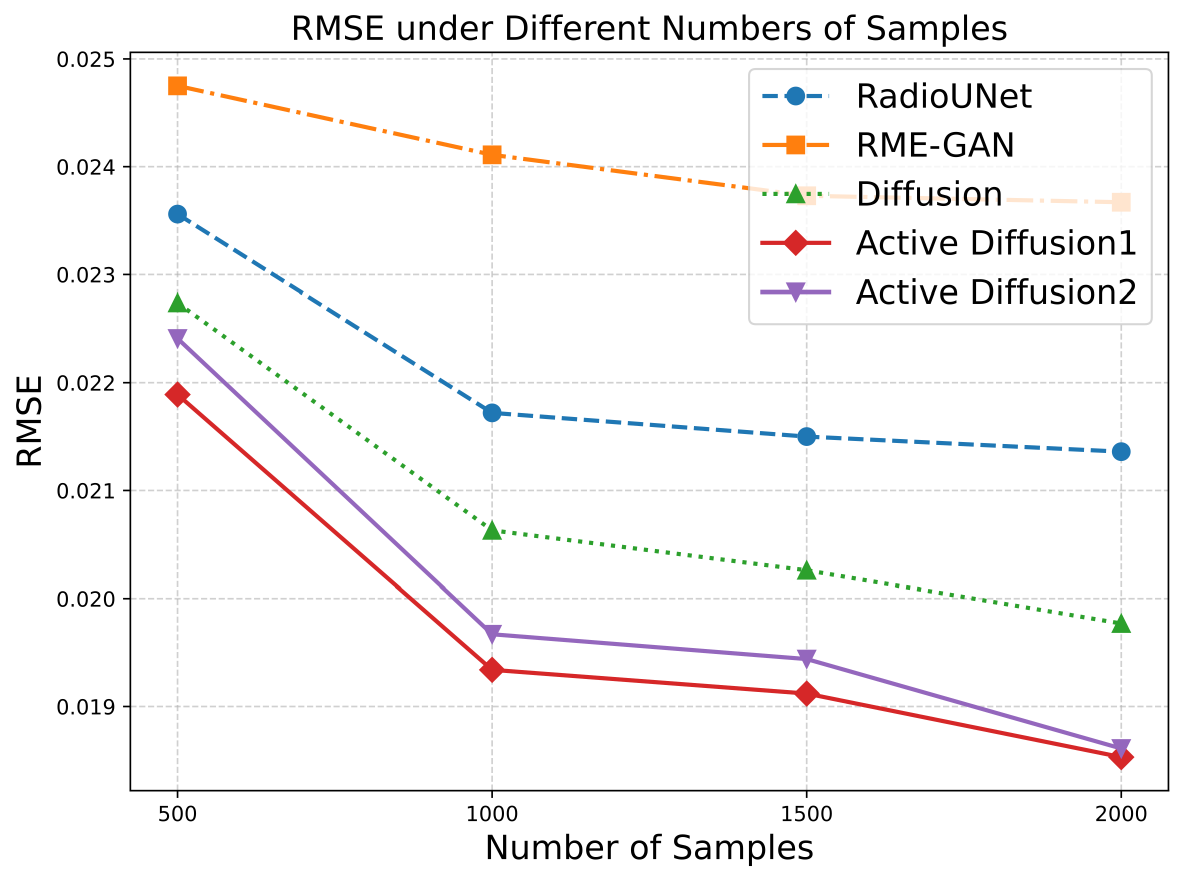}
    }
    \hfill
    \subfloat[\label{fig11_c}]{
        \includegraphics[width=0.31\textwidth]{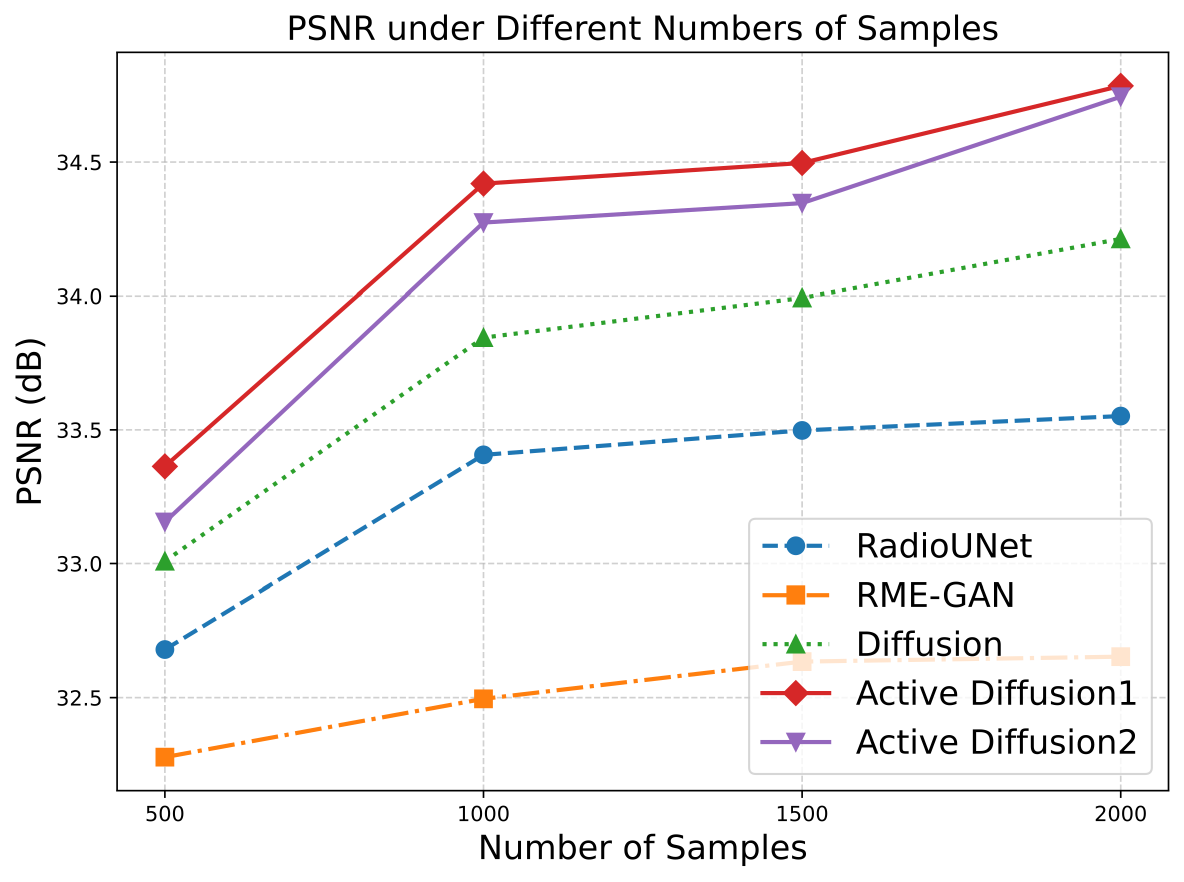}
    }
    \caption{Dynamic CGM. (A) NMSE. (B) RMSE. (C) PSNR.}
    \label{fig11}
\end{figure*}

Fig.~\ref{fig10} compares the static CGMs generated by different models with the ground-truth CGM when the number of sampling points is 1000, where the locations marked in red represent the sampling positions. RadioUNet often fails to preserve the sharp propagation-loss transitions at LOS/NLOS boundaries, leading to overly smoothed building blockage effects. Although RME-GAN improves edge sharpness, it still suffers from significant deviations in NLOS and transition areas. Conversely, diffusion-based approach effectively captures the complex spatial propagation of wireless signals, resolving both RadioUNet's over-smoothing and RME-GAN's inaccurate blockage reconstruction. However, its random sampling strategy may yield unrealistic distributions in obstructed regions. Our proposed method actively samples points with high predictive variance. These targeted observations provide crucial constraints that reduce reconstruction errors, thereby improving the spatial consistency between the constructed CGM and the true propagation field.

\begin{table}[!htbp]
    \begin{threeparttable}
        \caption{Construction Accuracy Comparison on the Dynamic CGM Dataset}
        \label{tab2}
        \centering
        \footnotesize
        \setlength{\tabcolsep}{5pt}
        \begin{tabular*}{\textwidth}{@{\extracolsep{\fill}}clccc@{}}
            \toprule
            Sampling Points & Method & NMSE & RMSE & PSNR (dB) \\
            \midrule
            \multirow{5}{*}{500}
            & RadioUNet & 0.00671 & 0.02356 & 32.67990 \\
            & RME-GAN & 0.00744 & 0.02475 & 32.27734 \\
            & Diffusion & 0.00644 & 0.02274 & 33.00844 \\
            & Active Diffusion 1 & \textbf{0.00601} & \textbf{0.02189} & \textbf{33.36332} \\
            & Active Diffusion 2 & \underline{0.00628} & \underline{0.02241} & \underline{33.15529} \\
            \addlinespace[1pt]
            \midrule
            \multirow{5}{*}{1000}
            & RadioUNet & 0.00572 & 0.02172 & 33.40645 \\
            & RME-GAN & 0.00708 & 0.02411 & 32.49545 \\
            & Diffusion & 0.00528 & 0.02063 & 33.84512 \\
            & Active Diffusion 1 & \textbf{0.00466} & \textbf{0.01934} & \textbf{34.41942} \\
            & Active Diffusion 2 & \underline{0.00483} & \underline{0.01967} & \underline{34.27374} \\
            \addlinespace[1pt]
            \midrule
            \multirow{5}{*}{1500}
            & RadioUNet & 0.00564 & 0.02150 & 33.49780 \\
            & RME-GAN & 0.00687 & 0.02373 & 32.63481 \\
            & Diffusion & 0.00506 & 0.02026 & 33.99313 \\
            & Active Diffusion 1 & \textbf{0.00450} & \textbf{0.01912} & \textbf{34.49663} \\
            & Active Diffusion 2 & \underline{0.00462} & \underline{0.01944} & \underline{34.34641} \\
            \addlinespace[1pt]
            \midrule
            \multirow{5}{*}{2000}
            & RadioUNet & 0.00558 & 0.02136 & 33.55159 \\
            & RME-GAN & 0.00677 & 0.02367 & 32.65327 \\
            & Diffusion & 0.00483 & 0.01977 & 34.21264 \\
            & Active Diffusion 1 & \textbf{0.00424} & \textbf{0.01853} & \textbf{34.78518} \\
            & Active Diffusion 2 & \underline{0.00426} & \underline{0.01861} & \underline{34.74442} \\
            \bottomrule
        \end{tabular*}
        \begin{tablenotes}
            \footnotesize
            \item[] \textbf{Note:} Bold and underlined values indicate the best and second-best performance, respectively.
        \end{tablenotes}
    \end{threeparttable}
\end{table}

\begin{figure}[!htbp]
\centering
\includegraphics[width=\columnwidth]{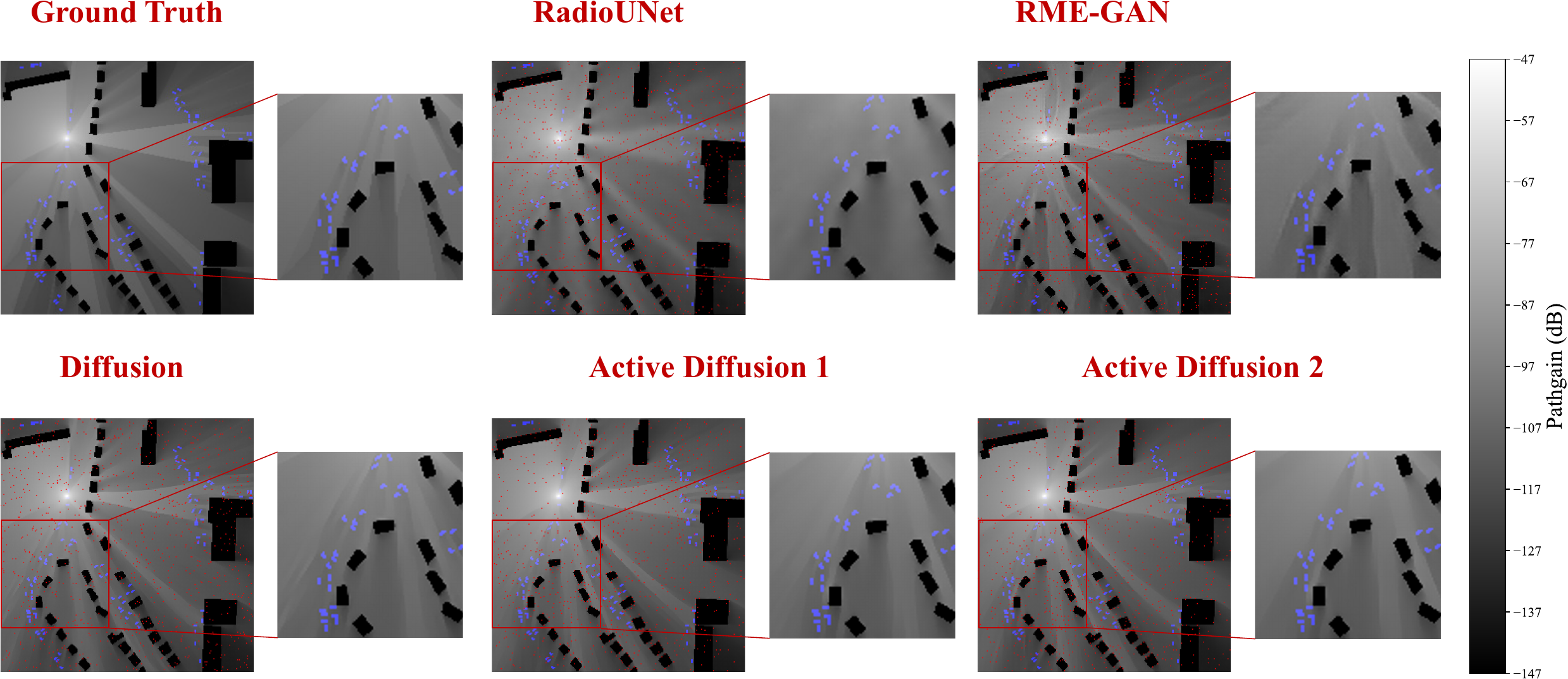}
\caption{Comparison of dynamic CGMs constructed by different models.}
\label{fig12}
\end{figure}

While performance trends on the dynamic CGM dataset align with the static results, the introduction of vehicles significantly complicates the propagation environment. This heightened complexity induces stronger uncertainty and severe local variations, leading to an overall drop in reconstruction accuracy. Despite this, the baseline diffusion model consistently outperforms RadioUNet and RME-GAN across all sampling budgets. The active learning mechanism further enhances accuracy: compared to random sampling, Active Diffusion 1 achieves NMSE reductions ranging from 6.68\% to 12.22\% and PSNR gains of 0.35~dB to 0.57~dB across the four sampling budgets. Active Diffusion 2 follows closely, yielding NMSE reductions of 2.48\% to 11.80\% and PSNR gains up to 0.53~dB, reaffirming the consistent superiority of the proposed framework.

Fig.~\ref{fig12} presents the dynamic CGMs constructed by different methods using 1000 sampling points, with the sampled locations highlighted in red. When vehicles are introduced as additional obstacles, RadioUNet still exhibits over-smoothing, and RME-GAN degrades more markedly. Although the diffusion model better captures the overall dynamic CGM distribution, it may still misrepresent local details in regions blocked by small obstacles such as vehicles. The two active-learning-based diffusion methods more effectively correct these local artifacts, thereby improving the spatial consistency between the reconstructed dynamic CGMs and the ground-truth field.

\section{Conclusion}\label{Conclusion}
This paper has addressed the CGM construction problem under sparse observations by proposing an active diffusion framework guided by epistemic uncertainty. The proposed method combines the generative reconstruction capability of diffusion models with Bayesian uncertainty quantification.
Furthermore, an uncertainty-aware sampling strategy is developed so that limited sampling points are preferentially allocated to regions with high uncertainty and reasonable spatial coverage. Experimental results on both static and dynamic CGM datasets demonstrate that our proposed method outperforms random-sampling baselines, thereby verifying the efficacy of leveraging epistemic uncertainty to guide the selection of active sampling points. Future work will extend active-learning-based framework to the construction of other types of channel knowledge maps.

\bibliography{reference}

@article{Zeng2021,
  author = {Yong Zeng and Xiaoli Xu},
  doi = {10.1109/MWC.001.2000327},
  issn = {15580687},
  issue = {3},
  journal = {IEEE Wireless Communications},
  month = {6},
  pages = {84-91},
  publisher = {Institute of Electrical and Electronics Engineers Inc.},
  title = {Toward Environment-Aware 6G Communications via Channel Knowledge Map},
  volume = {28},
  year = {2021}
}

@article{Zeng2024,
  author = {Yong Zeng and Junting Chen and Jie Xu and Di Wu and Xiaoli Xu and Shi Jin and Xiqi Gao and David Gesbert and Shuguang Cui and Rui Zhang},
  doi = {10.1109/COMST.2024.3364508},
  issn = {1553877X},
  issue = {3},
  journal = {IEEE Communications Surveys and Tutorials},
  pages = {1478-1519},
  publisher = {Institute of Electrical and Electronics Engineers Inc.},
  title = {A Tutorial on Environment-Aware Communications via Channel Knowledge Map for 6G},
  volume = {26},
  year = {2024}
}

@ARTICLE{9031749,
  author={Zou, Han and Chen, Chun-Lin and Li, Maoxun and Yang, Jianfei and Zhou, Yuxun and Xie, Lihua and Spanos, Costas J.},
  journal={IEEE Internet of Things Journal}, 
  title={Adversarial Learning-Enabled Automatic WiFi Indoor Radio Map Construction and Adaptation With Mobile Robot}, 
  year={2020},
  volume={7},
  number={8},
  pages={6946-6954},
  doi={10.1109/JIOT.2020.2979413}}

@ARTICLE{10025691,
  author={Li, Qiao and Liao, Xuewen and Li, Ang and Valaee, Shahrokh},
  journal={IEEE Transactions on Wireless Communications}, 
  title={Automatic Indoor Radio Map Construction and Localization via Multipath Fingerprint Extrapolation}, 
  year={2023},
  volume={22},
  number={9},
  pages={5814-5827},
  doi={10.1109/TWC.2023.3237359}}

@ARTICLE{8525324,
  author={Esrafilian, Omid and Gangula, Rajeev and Gesbert, David},
  journal={IEEE Internet of Things Journal}, 
  title={Learning to Communicate in UAV-Aided Wireless Networks: Map-Based Approaches}, 
  year={2019},
  volume={6},
  number={2},
  pages={1791-1802},
  doi={10.1109/JIOT.2018.2879682}}

@ARTICLE{9269485,
  author={Zhang, Shuowen and Zhang, Rui},
  journal={IEEE Transactions on Wireless Communications}, 
  title={Radio Map-Based 3D Path Planning for Cellular-Connected UAV}, 
  year={2021},
  volume={20},
  number={3},
  pages={1975-1989},
  doi={10.1109/TWC.2020.3037916}}

@ARTICLE{9354009,
  author={Zeng, Yong and Xu, Xiaoli and Jin, Shi and Zhang, Rui},
  journal={IEEE Transactions on Wireless Communications}, 
  title={Simultaneous Navigation and Radio Mapping for Cellular-Connected UAV With Deep Reinforcement Learning}, 
  year={2021},
  volume={20},
  number={7},
  pages={4205-4220},
  doi={10.1109/TWC.2021.3056573}}

@INPROCEEDINGS{10008712,
  author={Vankayala, Satya Kumar and Sharma, Kuldeep and Gollapudi, Sai Krishna Santosh and Singh, Sukhdeep and Qureshi, Nawab Mohammad Faseeh and Yoon, Seungil},
  booktitle={2022 IEEE Globecom Workshops (GC Wkshps)}, 
  title={Novel Localization Technique for Next Generation Base Stations using Radio maps}, 
  year={2022},
  volume={},
  number={},
  pages={783-788},
  doi={10.1109/GCWkshps56602.2022.10008712}}

@ARTICLE{10474197,
  author={Sallouha, Hazem and Sarkar, Shamik and Krijestorac, Enes and Cabric, Danijela},
  journal={IEEE Open Journal of Signal Processing}, 
  title={REM-U-Net: Deep Learning Based Agile REM Prediction With Energy-Efficient Cell-Free Use Case}, 
  year={2024},
  volume={5},
  number={},
  pages={750-765},
  doi={10.1109/OJSP.2024.3378591}}

@INPROCEEDINGS{8815467,
  author={Utkovski, Zoran and Agostini, Patrick and Frey, Matthias and Bjelakovic, Igor and Stanczak, Slawomir},
  booktitle={2019 IEEE 20th International Workshop on Signal Processing Advances in Wireless Communications (SPAWC)}, 
  title={Learning Radio Maps for Physical-Layer Security in the Radio Access}, 
  year={2019},
  volume={},
  number={},
  pages={1-5},
  doi={10.1109/SPAWC.2019.8815467}}

@INPROCEEDINGS{9053347,
  author={Levie, Ron and Yapar, Cagkan and Kutyniok, Gitta and Caire, Giuseppe},
  booktitle={ICASSP 2020 - 2020 IEEE International Conference on Acoustics, Speech and Signal Processing (ICASSP)}, 
  title={Pathloss Prediction using Deep Learning with Applications to Cellular Optimization and Efficient D2D Link Scheduling}, 
  year={2020},
  volume={},
  number={},
  pages={8678-8682},
  doi={10.1109/ICASSP40776.2020.9053347}}

@article{fu2025ckmdiff,
  title={CKMDiff: A generative diffusion model for CKM construction via inverse problems with learned priors},
  author={Fu, Shen and Zeng, Yong and Wu, Zijian and Wu, Di and Jin, Shi and Wang, Cheng-Xiang and Gao, Xiqi},
  journal={arXiv preprint arXiv:2504.17323},
  year={2025}
}

@article{Lu2008,
  author = {George Y Lu and David W Wong},
  doi = {10.1016/j.cageo.2007.07.010},
  issn = {0098-3004},
  issue = {9},
  journal = {Computers \& Geosciences},
  pages = {1044-1055},
  title = {An adaptive inverse-distance weighting spatial interpolation technique},
  volume = {34},
  url = {https://www.sciencedirect.com/science/article/pii/S0098300408000721},
  year = {2008}
}

@incollection{Bishop1995,
  author = {Bishop, Christopher M},
  isbn = {9780198538493},
  title = {Radial Basis Functions},
  booktitle = {Neural Networks for Pattern Recognition},
  publisher = {Oxford University Press},
  year = {1995},
  month = {11},
  doi = {10.1093/oso/9780198538493.003.0005},
  url = {https://doi.org/10.1093/oso/9780198538493.003.0005},
  eprint = {https://academic.oup.com/book/0/chapter/421885283/chapter-pdf/52331245/isbn-9780198538493-book-part-5.pdf},
}

@INPROCEEDINGS{1371308,
  author={van Beers, W.C.M. and Kleijnen, J.P.C.},
  booktitle={Proceedings of the 2004 Winter Simulation Conference, 2004.}, 
  title={Kriging interpolation in simulation: a survey}, 
  year={2004},
  volume={1},
  number={},
  pages={121},
  doi={10.1109/WSC.2004.1371308}}

@INPROCEEDINGS{7472907,
  author={Chouvardas, Symeon and Valentin, Stefan and Draief, Moez and Leconte, Mathieu},
  booktitle={2016 IEEE International Conference on Acoustics, Speech and Signal Processing (ICASSP)}, 
  title={A method to reconstruct coverage loss maps based on matrix completion and adaptive sampling}, 
  year={2016},
  volume={},
  number={},
  pages={6390-6394},
  doi={10.1109/ICASSP.2016.7472907}}

@ARTICLE{9523765,
  author={Teganya, Yves and Romero, Daniel},
  journal={IEEE Transactions on Wireless Communications}, 
  title={Deep Completion Autoencoders for Radio Map Estimation}, 
  year={2022},
  volume={21},
  number={3},
  pages={1710-1724},
  doi={10.1109/TWC.2021.3106154}}

@ARTICLE{10682525,
  author={Lee, Ju-Hyung and Molisch, Andreas F.},
  journal={IEEE Transactions on Wireless Communications}, 
  title={A Scalable and Generalizable Pathloss Map Prediction}, 
  year={2024},
  volume={23},
  number={11},
  pages={17793-17806},
  doi={10.1109/TWC.2024.3457431}}

@ARTICLE{9354041,
  author={Levie, Ron and Yapar, Cagkan and Kutyniok, Gitta and Caire, Giuseppe},
  journal={IEEE Transactions on Wireless Communications}, 
  title={RadioUNet: Fast Radio Map Estimation With Convolutional Neural Networks}, 
  year={2021},
  volume={20},
  number={6},
  pages={4001-4015},
  doi={10.1109/TWC.2021.3054977}}

@ARTICLE{11146461,
  author={Dai, Zhuoyin and Wu, Di and Xu, Xiaoli and Zeng, Yong},
  journal={IEEE Transactions on Vehicular Technology}, 
  title={Generating CKM Using Others' Data: Cross-AP CKM Inference With Deep Learning}, 
  year={2026},
  volume={75},
  number={2},
  pages={3360-3365},
  doi={10.1109/TVT.2025.3604862}}

@INPROCEEDINGS{11174709,
  author={Li, Yuxuan and Zhang, Cheng and Wang, Wen and Huang, Yongming},
  booktitle={2025 IEEE 101st Vehicular Technology Conference (VTC2025-Spring)}, 
  title={RMTransformer: Accurate Radio Map Construction and Coverage Prediction}, 
  year={2025},
  volume={},
  number={},
  pages={1-5},
  doi={10.1109/VTC2025-Spring65109.2025.11174709}}

@ARTICLE{10682510,
  author={Li, Xiaojie and Zhang, Songyang and Li, Hang and Li, Xiaoyang and Xu, Lexi and Xu, Haigao and Mei, Hui and Zhu, Guangxu and Qi, Nan and Xiao, Ming},
  journal={IEEE Transactions on Wireless Communications}, 
  title={RadioGAT: A Joint Model-Based and Data-Driven Framework for Multi-Band Radiomap Reconstruction via Graph Attention Networks}, 
  year={2024},
  volume={23},
  number={11},
  pages={17777-17792},
  doi={10.1109/TWC.2024.3457157}}

@ARTICLE{10078269,
  author={Chen, Guokai and Liu, Yongxiang and Zhang, Tao and Zhang, Jianzhao and Guo, Xiye and Yang, Jun},
  journal={IEEE Communications Letters}, 
  title={A Graph Neural Network Based Radio Map Construction Method for Urban Environment}, 
  year={2023},
  volume={27},
  number={5},
  pages={1327-1331},
  doi={10.1109/LCOMM.2023.3260272}}

@ARTICLE{8794603,
  author={Li, Zhuo and Cao, Jiannong and Wang, Hongwei and Zhao, Miao},
  journal={IEEE Journal on Selected Areas in Communications}, 
  title={Sparsely Self-Supervised Generative Adversarial Nets for Radio Frequency Estimation}, 
  year={2019},
  volume={37},
  number={11},
  pages={2428-2442},
  doi={10.1109/JSAC.2019.2933779}}

@ARTICLE{10130091,
  author={Zhang, Songyang and Wijesinghe, Achintha and Ding, Zhi},
  journal={IEEE Internet of Things Journal}, 
  title={RME-GAN: A Learning Framework for Radio Map Estimation Based on Conditional Generative Adversarial Network}, 
  year={2023},
  volume={10},
  number={20},
  pages={18016-18027},
  doi={10.1109/JIOT.2023.3278235}}

@ARTICLE{10764739,
  author={Wang, Xiucheng and Tao, Keda and Cheng, Nan and Yin, Zhisheng and Li, Zan and Zhang, Yuan and Shen, Xuemin},
  journal={IEEE Transactions on Cognitive Communications and Networking}, 
  title={RadioDiff: An Effective Generative Diffusion Model for Sampling-Free Dynamic Radio Map Construction}, 
  year={2025},
  volume={11},
  number={2},
  pages={738-750},
  doi={10.1109/TCCN.2024.3504489}}

@ARTICLE{11278649,
  author={Wang, Xiucheng and Zhang, Qiming and Cheng, Nan and Sun, Ruijin and Li, Zan and Cui, Shuguang and Shen, Xuemin},
  journal={IEEE Journal on Selected Areas in Communications}, 
  title={RadioDiff-k2: Helmholtz Equation Informed Generative Diffusion Model for Multi-Path Aware Radio Map Construction}, 
  year={2026},
  volume={44},
  number={},
  pages={2318-2333},
  doi={10.1109/JSAC.2025.3641105}}

@article{huang2026channel,
  title={Channel Knowledge Map Construction via Guided Flow Matching},
  author={Huang, Ziyu and Zeng, Yong and Fu, Shen and Xu, Xiaoli and Du, Hongyang},
  journal={arXiv preprint arXiv:2601.06156},
  year={2026}
}

@ARTICLE{10530520,
  author={Xu, Xiaoli and Zeng, Yong},
  journal={IEEE Transactions on Wireless Communications}, 
  title={How Much Data Is Needed for Channel Knowledge Map Construction?}, 
  year={2024},
  volume={23},
  number={10},
  pages={13011-13021},
  doi={10.1109/TWC.2024.3397964}}

@ARTICLE{10735108,
  author={Polyzos, Konstantinos D. and Sadeghi, Alireza and Ye, Wei and Sleder, Steven and Houssou, Kodjo and Calder, Jeff and Zhang, Zhi-Li and Giannakis, Georgios B.},
  journal={IEEE Transactions on Wireless Communications}, 
  title={Bayesian Active Learning for Sample Efficient {5G} Radio Map Reconstruction}, 
  year={2024},
  volume={23},
  number={12},
  pages={19382-19396},
  doi={10.1109/TWC.2024.3483112}}

@article{ho2020denoising,
  title={Denoising diffusion probabilistic models},
  author={Ho, Jonathan and Jain, Ajay and Abbeel, Pieter},
  journal={Advances in neural information processing systems},
  volume={33},
  pages={6840--6851},
  year={2020}
}

@article{Song2020DenoisingDI,
  title={Denoising Diffusion Implicit Models},
  author={Jiaming Song and Chenlin Meng and Stefano Ermon},
  journal={ArXiv},
  year={2020},
  volume={abs/2010.02502},
  url={https://api.semanticscholar.org/CorpusID:222140788}
}

@article{Rombach2021HighResolutionIS,
  title={High-Resolution Image Synthesis with Latent Diffusion Models},
  author={Robin Rombach and A. Blattmann and Dominik Lorenz and Patrick Esser and Bj{\"o}rn Ommer},
  journal={2022 IEEE/CVF Conference on Computer Vision and Pattern Recognition (CVPR)},
  year={2021},
  pages={10674-10685},
  url={https://api.semanticscholar.org/CorpusID:245335280}
}

@article{chan2024estimating,
  title={Estimating epistemic and aleatoric uncertainty with a single model},
  author={Chan, Matthew A and Molina, Maria J and Metzler, Christopher A},
  journal={Advances in Neural Information Processing Systems},
  volume={37},
  pages={109845--109870},
  year={2024}
}

@inproceedings{blundell2015weight,
  title={Weight uncertainty in neural network},
  author={Blundell, Charles and Cornebise, Julien and Kavukcuoglu, Koray and Wierstra, Daan},
  booktitle={International conference on machine learning},
  pages={1613--1622},
  year={2015},
  organization={PMLR}
}

@inproceedings{hernandez2015probabilistic,
  title={Probabilistic backpropagation for scalable learning of bayesian neural networks},
  author={Hern{\'a}ndez-Lobato, Jos{\'e} Miguel and Adams, Ryan},
  booktitle={International conference on machine learning},
  pages={1861--1869},
  year={2015},
  organization={PMLR}
}

@inproceedings{chen2014stochastic,
  title={Stochastic gradient hamiltonian monte carlo},
  author={Chen, Tianqi and Fox, Emily and Guestrin, Carlos},
  booktitle={International conference on machine learning},
  pages={1683--1691},
  year={2014},
  organization={PMLR}
}

@article{lakshminarayanan2017simple,
  title={Simple and scalable predictive uncertainty estimation using deep ensembles},
  author={Lakshminarayanan, Balaji and Pritzel, Alexander and Blundell, Charles},
  journal={Advances in neural information processing systems},
  volume={30},
  year={2017}
}

@inproceedings{gal2016dropout,
  title={Dropout as a bayesian approximation: Representing model uncertainty in deep learning},
  author={Gal, Yarin and Ghahramani, Zoubin},
  booktitle={international conference on machine learning},
  pages={1050--1059},
  year={2016},
  organization={PMLR}
}

@inproceedings{ritter2018scalable,
  title={A scalable laplace approximation for neural networks},
  author={Ritter, Hippolyt and Botev, Aleksandar and Barber, David},
  booktitle={International conference on learning representations},
  year={2018}
}

@article{daxberger2021laplace,
  title={Laplace redux-effortless bayesian deep learning},
  author={Daxberger, Erik and Kristiadi, Agustinus and Immer, Alexander and Eschenhagen, Runa and Bauer, Matthias and Hennig, Philipp},
  journal={Advances in neural information processing systems},
  volume={34},
  pages={20089--20103},
  year={2021}
}

@inproceedings{kristiadi2020being,
  title={Being bayesian, even just a bit, fixes overconfidence in relu networks},
  author={Kristiadi, Agustinus and Hein, Matthias and Hennig, Philipp},
  booktitle={International conference on machine learning},
  pages={5436--5446},
  year={2020},
  organization={PMLR}
}

@misc{0gtx6v3022,
  doi = {10.21227/0gtx-6v30},
  url = {https://dx.doi.org/10.21227/0gtx-6v30},
  author = {Cagkan Yapar and Ron Levie and Gitta Kutyniok and Giuseppe Caire},
  publisher = {IEEE Dataport},
  title = {Dataset of Pathloss and ToA Radio Maps with Localization Application},
  year = {2022}
}
\bibliographystyle{plain}
\end{document}